%% file: main.tex
\documentclass{llncs}

\usepackage[FIGBOTCAP,TABBOTCAP,center,nooneline]{subfigure}
\usepackage[dvips]{rotating}
\usepackage{RSalgorithmic}
\usepackage{algorithm}
\usepackage{amscd}
\usepackage{amsfonts}
\usepackage{amsmath}
\usepackage{amssymb}
\usepackage{booktabs}
\usepackage{color}
\usepackage{datetime}
\usepackage{epsfig}
\usepackage{graphicx}
\usepackage{latexsym}
\usepackage{marginnote}
\usepackage{mathtools}
\usepackage{multirow}
\usepackage{paralist}
\usepackage{placeins}
\usepackage{rotate} 
\usepackage{subfig}
\usepackage{tikz}
\usepackage{times} 
\usepackage{url}
\usepackage{varwidth}
\usepackage{verbatim} 
\usepackage{wrapfig}
\usepackage{xcolor,colortbl}
\usepackage{xspace}

\input{rs_macros_colors}

\input{rs_macros_dl}
\input{rs_macros_general}

\input{rs_macros_mc}

\input{rs_macros_optsmt}

\input{rs_macros_sat}

\input{rs_macros_smt}

\input{rs_macros_specific}
\input{pt_macros}

 \renewcommand{\ignoreinshort}[1]{}
 
\newcommand{\PTTODO}[1]{\noindent{\bf \textcolor{blue}{{{[PT TODO]:} #1]}}}}
\renewcommand{\RSTODO}[1]{\noindent{\bf \textcolor{blue}{{{[RS TODO]:} #1}}}}
\renewcommand{\TODO}[1]{\noindent{\bf \textcolor{blue}{{{[TODO]:} #1}}}}
\begin{document}

\pagestyle{plain}
\pagenumbering{roman}

\pagestyle{plain}
\pagenumbering{arabic}

\title{%
% On the Benefits of Enhancing 
% Optimization Modulo Theories with
% Sorting Networks for MaxSMT
% 
% \thanks{This work is supported by %\ignoreinlong{SRC}
% \ignoreinshort{{Semiconductor Research Corporation} (SRC)}
%   under
% GRC Research Project 2012-TJ-2266 WOLF.
% }
%
\ignore{
Enhancing 
Optimization Modulo Theories \\with
Sorting Networks %for MaxSMT
{\\ \small or \\}
On the Benefits of 
Enhancing 
Optimization Modulo Theories with
Sorting Networks %for MaxSMT
{\\ \small or \\}
}
On Optimization Modulo Theories, MaxSMT and Sorting Networks
}

\author{
Roberto Sebastiani 
\and 
Patrick Trentin
}

\institute{%
DISI, University of Trento, Italy%
}

\maketitle
\ignoreinshort{
\large
\begin{center}
\noi
{\em Latest update: \today}
\end{center}
}
 \begin{abstract}
 \input{abstract}
 \end{abstract}

%%%%%%%%%%%%%%%%%%%%%%%%%%%%%%%%%%%%%%%%%%%%%%%%%%%%%%%%%%%%%
%%% 
%%%%%%%%%%%%%%%%%%%%%%%%%%%%%%%%%%%%%%%%%%%%%%%%%%%%%%%%%%%%%
\section{Introduction}
\label{sec:intro}
\input{intro}

%%%%%%%%%%%%%%%%%%%%%%%%%%%%%%%%%%%%%%%%%%%%%%%%%%%%%%%%%%%%%
%%% 
%%%%%%%%%%%%%%%%%%%%%%%%%%%%%%%%%%%%%%%%%%%%%%%%%%%%%%%%%%%%%

%%%%%%%%%%%%%%%%%%%%%%%%%%%%%%%%%%%%%%%%%%%%%%%%%%%%%%%%%%%%%
%%% 
%%%%%%%%%%%%%%%%%%%%%%%%%%%%%%%%%%%%%%%%%%%%%%%%%%%%%%%%%%%%%
%\section{Background and State of the Art}
\section{Background}
\label{sec:background}
\input{background}

%\subsubsection*{Related Work.}
%\label{sec:related}
%\input{related}

%\newpage

%%%%%%%%%%%%%%%%%%%%%%%%%%%%%%%%%%%%%%%%%%%%%%%%%%%%%%%%%%%%%
%%% 
%%%%%%%%%%%%%%%%%%%%%%%%%%%%%%%%%%%%%%%%%%%%%%%%%%%%%%%%%%%%%
\section{Problems with OMT-based Approaches}
\label{sec:problem}
\input{problem}

%\newpage
%%%%%%%%%%%%%%%%%%%%%%%%%%%%%%%%%%%%%%%%%%%%%%%%%%%%%%%%%%%%%
%%% 
%%%%%%%%%%%%%%%%%%%%%%%%%%%%%%%%%%%%%%%%%%%%%%%%%%%%%%%%%%%%%
\section{Combining OMT with Sorting Networks}
\label{sec:combining}
\input{combining}
\input{combining_trail}

%%%%%%%%%%%%%%%%%%%%%%%%%%%%%%%%%%%%%%%%%%%%%%%%%%%%%%%%%%%%%
%%% 
%%%%%%%%%%%%%%%%%%%%%%%%%%%%%%%%%%%%%%%%%%%%%%%%%%%%%%%%%%%%%
%\newpage
\section{Experimental Evaluation}
\label{sec:expeval}
\input{expeval}
\subsection{Problems suitable for MaxSAT-based approaches}
\label{sec:expeval_suitable}
\subsubsection{Test Set \#1: CGMs with lexicographic PB optimization.}
\label{sec:expeval_suitable_lexicographic}
\input{expeval_lexicographic}
\subsubsection{Test Set \#2: CGMs with weight-1 PB optimization.}
\label{sec:expeval_suitable_puremaxsmt}
\input{expeval_puremaxsmt}
\input{expeval_ll}
\subsection{Problems unsuitable for MaxSAT-based approaches}
\input{unsuitable.tex}

\label{sec:expeval_unsuitable}
\subsubsection{Test Set \#3: CGMs with max-min PB optimization.}
\label{sec:expeval_unsuitable_maxmin}
\input{expeval_maxmin}

\subsubsection{Test Set \#4: LMT with mixed complex objective functions.}
\label{sec:expeval_unsuitable_mixed}
\input{expeval_mixed_lmt}
\subsubsection*{Discussion.}
\label{sec:expeval_discussion}
\input{expeval_discussion}

%%%%%%%%%%%%%%%%%%%%%%%%%%%%%%%%%%%%%%%%%%%%%%%%%%%%%%%%%%%%%
%%% 
%%%%%%%%%%%%%%%%%%%%%%%%%%%%%%%%%%%%%%%%%%%%%%%%%%%%%%%%%%%%%
\section{Conclusion and Future Work}
\label{sec:concl}
\input{concl}

\newpage
\FloatBarrier
\pagenumbering{roman}
\bibliographystyle{abbrv} 
\bibliography{rs_refs,rs_ownrefs,rs_specific,sathandbook,pt_refs}

\end{document}

%% file: rs_macros_colors.tex
\newcommand{\blue}[1]{\textcolor{blue}{#1}}

\newcommand{\green}[1]{\textcolor{darkgreen}{#1}}

%% file: rs_macros_general.tex
\input{rgb}

\newcommand{\resp}[1]{[resp. #1]}

\newcommand{\TODO}[1]{{}}

\newcommand{\ignore}[1]{}
\newcommand{\todo}[1] {}
\newcommand{\RSCHANGE}[1]{\textcolor{darkgreen}{#1}}

\newcommand{\RSTODO}[1]{{\bf \textcolor{darkgreen}{{\fbox{RS TODO:} #1}}}}
\renewcommand{\RSTODO}[1]{}
 \newcommand{\ignoreinshort}[1]{}
 
\def\makenewenumerate#1#2{%
\newcounter{cnt#1}
\newenvironment{#1}%
{\begin{list}{\makebox[0pt][r]{#2}}%
{\setlength{\itemsep}{0pt}% 
 \setlength{\parsep}{.2em}%
 \setlength{\leftmargin}{1.5em}%
 \setlength{\labelwidth}{.4em}%
 \usecounter{cnt#1}}}
{\end{list}}}

\makenewenumerate{myenumerate}{\arabic{cntmyenumerate}.}

\makenewenumerate{renumerate}{\rm(\roman{cntrenumerate})}
\makenewenumerate{renumerateprime}{\rm(\roman{cntrenumerateprime}$'$)}
\makenewenumerate{renumeratesecond}{\rm(\roman{cntrenumeratesecond}$''$)}

\makenewenumerate{aenumerate}{({\it\alph{cntaenumerate}})}
\makenewenumerate{aenumerateprime}{({\it\alph{cntaenumerateprime}$'$})}
\makenewenumerate{aenumeratesecond}{({\it\alph{cntaenumeratesecond}$''$})}

\def\newplaintheorem#1#2{%
\newtheorem{#1plain}{#2}[section]%
\newenvironment{#1}{\begin{#1plain}\rm }{\end{#1plain}}}

\newplaintheorem{notation}{Notation}
\newplaintheorem{cexample}{Counter-Example}

\newcommand{\sref}[1]{\S{}\ref{#1}}
\newcommand{\noi}{\noindent}

\newcommand{\tuple}[1]{\ensuremath{\langle{#1}\rangle}\xspace}
\newcommand{\set}[1]{\ensuremath{\{{#1}\}}\xspace}
\newcommand{\imp}{\ensuremath{\rightarrow}\xspace}
\newcommand{\limp}{\ensuremath{\leftarrow}\xspace}

\newcommand{\defas}{\ensuremath{\stackrel{\text{\tiny def}}{=}}\xspace}

\newcommand{\pos}{\phantom{\neg}}

\newcommand\calc{\ensuremath{\mathcal{C}}\xspace}

\renewcommand{\RSCHANGE}[1]{\textcolor{darkgreen}{{#1}}}

%% file: rgb.tex
\definecolor {snow}                {rgb}{1.00,0.98,0.98}
\definecolor {ghostwhite}          {rgb}{0.97,0.97,1.00}
\definecolor {whitesmoke}          {rgb}{0.96,0.96,0.96}
\definecolor {gainsboro}           {rgb}{0.86,0.86,0.86}
\definecolor {floralwhite}         {rgb}{1.00,0.98,0.94}
\definecolor {oldlace}             {rgb}{0.99,0.96,0.90}
\definecolor {linen}               {rgb}{0.98,0.94,0.90}
\definecolor {antiquewhite}        {rgb}{0.98,0.92,0.84}
\definecolor {papayawhip}          {rgb}{1.00,0.94,0.84}
\definecolor {blanchedalmond}      {rgb}{1.00,0.92,0.80}
\definecolor {bisque}              {rgb}{1.00,0.89,0.77}
\definecolor {peachpuff}           {rgb}{1.00,0.85,0.73}
\definecolor {navajowhite}         {rgb}{1.00,0.87,0.68}
\definecolor {moccasin}            {rgb}{1.00,0.89,0.71}
\definecolor {cornsilk}            {rgb}{1.00,0.97,0.86}
\definecolor {ivory}               {rgb}{1.00,1.00,0.94}
\definecolor {lemonchiffon}        {rgb}{1.00,0.98,0.80}
\definecolor {seashell}            {rgb}{1.00,0.96,0.93}
\definecolor {honeydew}            {rgb}{0.94,1.00,0.94}
\definecolor {mintcream}           {rgb}{0.96,1.00,0.98}
\definecolor {azure}               {rgb}{0.94,1.00,1.00}
\definecolor {aliceblue}           {rgb}{0.94,0.97,1.00}
\definecolor {lavender}            {rgb}{0.90,0.90,0.98}
\definecolor {lavenderblush}       {rgb}{1.00,0.94,0.96}
\definecolor {mistyrose}           {rgb}{1.00,0.89,0.88}
\definecolor {white}               {rgb}{1.00,1.00,1.00}
\definecolor {black}               {rgb}{0.00,0.00,0.00}
\definecolor {darkslategray}       {rgb}{0.18,0.31,0.31}
\definecolor {dimgray}             {rgb}{0.41,0.41,0.41}
\definecolor {slategray}           {rgb}{0.44,0.50,0.56}
\definecolor {lightslategray}      {rgb}{0.47,0.53,0.60}
\definecolor {gray}                {rgb}{0.75,0.75,0.75}
\definecolor {lightgrey}           {rgb}{0.83,0.83,0.83}
\definecolor {midnightblue}        {rgb}{0.10,0.10,0.44}
\definecolor {navy}                {rgb}{0.00,0.00,0.50}
\definecolor {cornflowerblue}      {rgb}{0.39,0.58,0.93}
\definecolor {darkslateblue}       {rgb}{0.28,0.24,0.55}
\definecolor {slateblue}           {rgb}{0.42,0.35,0.80}
\definecolor {mediumslateblue}     {rgb}{0.48,0.41,0.93}
\definecolor {lightslateblue}      {rgb}{0.52,0.44,1.00}
\definecolor {mediumblue}          {rgb}{0.00,0.00,0.80}
\definecolor {royalblue}           {rgb}{0.25,0.41,0.88}
\definecolor {blue}                {rgb}{0.00,0.00,1.00}
\definecolor {dodgerblue}          {rgb}{0.12,0.56,1.00}
\definecolor {deepskyblue}         {rgb}{0.00,0.75,1.00}
\definecolor {skyblue}             {rgb}{0.53,0.81,0.92}
\definecolor {lightskyblue}        {rgb}{0.53,0.81,0.98}
\definecolor {steelblue}           {rgb}{0.27,0.51,0.71}
\definecolor {lightsteelblue}      {rgb}{0.69,0.77,0.87}
\definecolor {lightblue}           {rgb}{0.68,0.85,0.90}
\definecolor {powderblue}          {rgb}{0.69,0.88,0.90}
\definecolor {paleturquoise}       {rgb}{0.69,0.93,0.93}
\definecolor {darkturquoise}       {rgb}{0.00,0.81,0.82}
\definecolor {mediumturquoise}     {rgb}{0.28,0.82,0.80}
\definecolor {turquoise}           {rgb}{0.25,0.88,0.82}
\definecolor {cyan}                {rgb}{0.00,1.00,1.00}
\definecolor {lightcyan}           {rgb}{0.88,1.00,1.00}
\definecolor {cadetblue}           {rgb}{0.37,0.62,0.63}
\definecolor {mediumaquamarine}    {rgb}{0.40,0.80,0.67}
\definecolor {aquamarine}          {rgb}{0.50,1.00,0.83}
\definecolor {darkgreen}           {rgb}{0.00,0.39,0.00}
\definecolor {darkolivegreen}      {rgb}{0.33,0.42,0.18}
\definecolor {darkseagreen}        {rgb}{0.56,0.74,0.56}
\definecolor {seagreen}            {rgb}{0.18,0.55,0.34}
\definecolor {mediumseagreen}      {rgb}{0.24,0.70,0.44}
\definecolor {lightseagreen}       {rgb}{0.13,0.70,0.67}
\definecolor {palegreen}           {rgb}{0.60,0.98,0.60}
\definecolor {springgreen}         {rgb}{0.00,1.00,0.50}
\definecolor {lawngreen}           {rgb}{0.49,0.99,0.00}
\definecolor {green}               {rgb}{0.00,1.00,0.00}
\definecolor {chartreuse}          {rgb}{0.50,1.00,0.00}
\definecolor {mediumspringgreen}   {rgb}{0.00,0.98,0.60}
\definecolor {greenyellow}         {rgb}{0.68,1.00,0.18}
\definecolor {limegreen}           {rgb}{0.20,0.80,0.20}
\definecolor {yellowgreen}         {rgb}{0.60,0.80,0.20}
\definecolor {forestgreen}         {rgb}{0.13,0.55,0.13}
\definecolor {olivedrab}           {rgb}{0.42,0.56,0.14}
\definecolor {darkkhaki}           {rgb}{0.74,0.72,0.42}
\definecolor {khaki}               {rgb}{0.94,0.90,0.55}
\definecolor {palegoldenrod}       {rgb}{0.93,0.91,0.67}
\definecolor {lightgoldenrodyellow} {rgb}{0.98,0.98,0.82}
\definecolor {lightyellow}         {rgb}{1.00,1.00,0.88}
\definecolor {yellow}              {rgb}{1.00,1.00,0.00}
\definecolor {gold}                {rgb}{1.00,0.84,0.00}
\definecolor {lightgoldenrod}      {rgb}{0.93,0.87,0.51}
\definecolor {goldenrod}           {rgb}{0.85,0.65,0.13}
\definecolor {darkgoldenrod}       {rgb}{0.72,0.53,0.04}
\definecolor {rosybrown}           {rgb}{0.74,0.56,0.56}
\definecolor {indianred}           {rgb}{0.80,0.36,0.36}
\definecolor {saddlebrown}         {rgb}{0.55,0.27,0.07}
\definecolor {sienna}              {rgb}{0.63,0.32,0.18}
\definecolor {peru}                {rgb}{0.80,0.52,0.25}
\definecolor {burlywood}           {rgb}{0.87,0.72,0.53}
\definecolor {beige}               {rgb}{0.96,0.96,0.86}
\definecolor {wheat}               {rgb}{0.96,0.87,0.70}
\definecolor {sandybrown}          {rgb}{0.96,0.64,0.38}
\definecolor {tan}                 {rgb}{0.82,0.71,0.55}
\definecolor {chocolate}           {rgb}{0.82,0.41,0.12}
\definecolor {firebrick}           {rgb}{0.70,0.13,0.13}
\definecolor {brown}               {rgb}{0.65,0.16,0.16}
\definecolor {darksalmon}          {rgb}{0.91,0.59,0.48}
\definecolor {salmon}              {rgb}{0.98,0.50,0.45}
\definecolor {lightsalmon}         {rgb}{1.00,0.63,0.48}
\definecolor {orange}              {rgb}{1.00,0.65,0.00}
\definecolor {darkorange}          {rgb}{1.00,0.55,0.00}
\definecolor {coral}               {rgb}{1.00,0.50,0.31}
\definecolor {lightcoral}          {rgb}{0.94,0.50,0.50}
\definecolor {tomato}              {rgb}{1.00,0.39,0.28}
\definecolor {orangered}           {rgb}{1.00,0.27,0.00}
\definecolor {red}                 {rgb}{1.00,0.00,0.00}
\definecolor {hotpink}             {rgb}{1.00,0.41,0.71}
\definecolor {deeppink}            {rgb}{1.00,0.08,0.58}
\definecolor {pink}                {rgb}{1.00,0.75,0.80}
\definecolor {lightpink}           {rgb}{1.00,0.71,0.76}
\definecolor {palevioletred}       {rgb}{0.86,0.44,0.58}
\definecolor {maroon}              {rgb}{0.69,0.19,0.38}
\definecolor {mediumvioletred}     {rgb}{0.78,0.08,0.52}
\definecolor {violetred}           {rgb}{0.82,0.13,0.56}
\definecolor {magenta}             {rgb}{1.00,0.00,1.00}
\definecolor {violet}              {rgb}{0.93,0.51,0.93}
\definecolor {plum}                {rgb}{0.87,0.63,0.87}
\definecolor {orchid}              {rgb}{0.85,0.44,0.84}
\definecolor {mediumorchid}        {rgb}{0.73,0.33,0.83}
\definecolor {darkorchid}          {rgb}{0.60,0.20,0.80}
\definecolor {darkviolet}          {rgb}{0.58,0.00,0.83}
\definecolor {blueviolet}          {rgb}{0.54,0.17,0.89}
\definecolor {purple}              {rgb}{0.63,0.13,0.94}
\definecolor {mediumpurple}        {rgb}{0.58,0.44,0.86}
\definecolor {thistle}             {rgb}{0.85,0.75,0.85}
\definecolor {snow2}               {rgb}{0.93,0.91,0.91}
\definecolor {snow3}               {rgb}{0.80,0.79,0.79}
\definecolor {snow4}               {rgb}{0.55,0.54,0.54}
\definecolor {seashell2}           {rgb}{0.93,0.90,0.87}
\definecolor {seashell3}           {rgb}{0.80,0.77,0.75}
\definecolor {seashell4}           {rgb}{0.55,0.53,0.51}
\definecolor {antiquewhite1}       {rgb}{1.00,0.94,0.86}
\definecolor {antiquewhite2}       {rgb}{0.93,0.87,0.80}
\definecolor {antiquewhite3}       {rgb}{0.80,0.75,0.69}
\definecolor {antiquewhite4}       {rgb}{0.55,0.51,0.47}
\definecolor {bisque2}             {rgb}{0.93,0.84,0.72}
\definecolor {bisque3}             {rgb}{0.80,0.72,0.62}
\definecolor {bisque4}             {rgb}{0.55,0.49,0.42}
\definecolor {peachpuff2}          {rgb}{0.93,0.80,0.68}
\definecolor {peachpuff3}          {rgb}{0.80,0.69,0.58}
\definecolor {peachpuff4}          {rgb}{0.55,0.47,0.40}
\definecolor {navajowhite2}        {rgb}{0.93,0.81,0.63}
\definecolor {navajowhite3}        {rgb}{0.80,0.70,0.55}
\definecolor {navajowhite4}        {rgb}{0.55,0.47,0.37}
\definecolor {lemonchiffon2}       {rgb}{0.93,0.91,0.75}
\definecolor {lemonchiffon3}       {rgb}{0.80,0.79,0.65}
\definecolor {lemonchiffon4}       {rgb}{0.55,0.54,0.44}
\definecolor {cornsilk2}           {rgb}{0.93,0.91,0.80}
\definecolor {cornsilk3}           {rgb}{0.80,0.78,0.69}
\definecolor {cornsilk4}           {rgb}{0.55,0.53,0.47}
\definecolor {ivory2}              {rgb}{0.93,0.93,0.88}
\definecolor {ivory3}              {rgb}{0.80,0.80,0.76}
\definecolor {ivory4}              {rgb}{0.55,0.55,0.51}
\definecolor {honeydew2}           {rgb}{0.88,0.93,0.88}
\definecolor {honeydew3}           {rgb}{0.76,0.80,0.76}
\definecolor {honeydew4}           {rgb}{0.51,0.55,0.51}
\definecolor {lavenderblush2}      {rgb}{0.93,0.88,0.90}
\definecolor {lavenderblush3}      {rgb}{0.80,0.76,0.77}
\definecolor {lavenderblush4}      {rgb}{0.55,0.51,0.53}
\definecolor {mistyrose2}          {rgb}{0.93,0.84,0.82}
\definecolor {mistyrose3}          {rgb}{0.80,0.72,0.71}
\definecolor {mistyrose4}          {rgb}{0.55,0.49,0.48}
\definecolor {azure2}              {rgb}{0.88,0.93,0.93}
\definecolor {azure3}              {rgb}{0.76,0.80,0.80}
\definecolor {azure4}              {rgb}{0.51,0.55,0.55}
\definecolor {slateblue1}          {rgb}{0.51,0.44,1.00}
\definecolor {slateblue2}          {rgb}{0.48,0.40,0.93}
\definecolor {slateblue3}          {rgb}{0.41,0.35,0.80}
\definecolor {slateblue4}          {rgb}{0.28,0.24,0.55}
\definecolor {royalblue1}          {rgb}{0.28,0.46,1.00}
\definecolor {royalblue2}          {rgb}{0.26,0.43,0.93}
\definecolor {royalblue3}          {rgb}{0.23,0.37,0.80}
\definecolor {royalblue4}          {rgb}{0.15,0.25,0.55}
\definecolor {blue2}               {rgb}{0.00,0.00,0.93}
\definecolor {blue4}               {rgb}{0.00,0.00,0.55}
\definecolor {dodgerblue2}         {rgb}{0.11,0.53,0.93}
\definecolor {dodgerblue3}         {rgb}{0.09,0.45,0.80}
\definecolor {dodgerblue4}         {rgb}{0.06,0.31,0.55}
\definecolor {steelblue1}          {rgb}{0.39,0.72,1.00}
\definecolor {steelblue2}          {rgb}{0.36,0.67,0.93}
\definecolor {steelblue3}          {rgb}{0.31,0.58,0.80}
\definecolor {steelblue4}          {rgb}{0.21,0.39,0.55}
\definecolor {deepskyblue2}        {rgb}{0.00,0.70,0.93}
\definecolor {deepskyblue3}        {rgb}{0.00,0.60,0.80}
\definecolor {deepskyblue4}        {rgb}{0.00,0.41,0.55}
\definecolor {skyblue1}            {rgb}{0.53,0.81,1.00}
\definecolor {skyblue2}            {rgb}{0.49,0.75,0.93}
\definecolor {skyblue3}            {rgb}{0.42,0.65,0.80}
\definecolor {skyblue4}            {rgb}{0.29,0.44,0.55}
\definecolor {lightskyblue1}       {rgb}{0.69,0.89,1.00}
\definecolor {lightskyblue2}       {rgb}{0.64,0.83,0.93}
\definecolor {lightskyblue3}       {rgb}{0.55,0.71,0.80}
\definecolor {lightskyblue4}       {rgb}{0.38,0.48,0.55}
\definecolor {slategray1}          {rgb}{0.78,0.89,1.00}
\definecolor {slategray2}          {rgb}{0.73,0.83,0.93}
\definecolor {slategray3}          {rgb}{0.62,0.71,0.80}
\definecolor {slategray4}          {rgb}{0.42,0.48,0.55}
\definecolor {lightsteelblue1}     {rgb}{0.79,0.88,1.00}
\definecolor {lightsteelblue2}     {rgb}{0.74,0.82,0.93}
\definecolor {lightsteelblue3}     {rgb}{0.64,0.71,0.80}
\definecolor {lightsteelblue4}     {rgb}{0.43,0.48,0.55}
\definecolor {lightblue1}          {rgb}{0.75,0.94,1.00}
\definecolor {lightblue2}          {rgb}{0.70,0.87,0.93}
\definecolor {lightblue3}          {rgb}{0.60,0.75,0.80}
\definecolor {lightblue4}          {rgb}{0.41,0.51,0.55}
\definecolor {lightcyan2}          {rgb}{0.82,0.93,0.93}
\definecolor {lightcyan3}          {rgb}{0.71,0.80,0.80}
\definecolor {lightcyan4}          {rgb}{0.48,0.55,0.55}
\definecolor {paleturquoise1}      {rgb}{0.73,1.00,1.00}
\definecolor {paleturquoise2}      {rgb}{0.68,0.93,0.93}
\definecolor {paleturquoise3}      {rgb}{0.59,0.80,0.80}
\definecolor {paleturquoise4}      {rgb}{0.40,0.55,0.55}
\definecolor {cadetblue1}          {rgb}{0.60,0.96,1.00}
\definecolor {cadetblue2}          {rgb}{0.56,0.90,0.93}
\definecolor {cadetblue3}          {rgb}{0.48,0.77,0.80}
\definecolor {cadetblue4}          {rgb}{0.33,0.53,0.55}
\definecolor {turquoise1}          {rgb}{0.00,0.96,1.00}
\definecolor {turquoise2}          {rgb}{0.00,0.90,0.93}
\definecolor {turquoise3}          {rgb}{0.00,0.77,0.80}
\definecolor {turquoise4}          {rgb}{0.00,0.53,0.55}
\definecolor {cyan2}               {rgb}{0.00,0.93,0.93}
\definecolor {cyan3}               {rgb}{0.00,0.80,0.80}
\definecolor {cyan4}               {rgb}{0.00,0.55,0.55}
\definecolor {darkslategray1}      {rgb}{0.59,1.00,1.00}
\definecolor {darkslategray2}      {rgb}{0.55,0.93,0.93}
\definecolor {darkslategray3}      {rgb}{0.47,0.80,0.80}
\definecolor {darkslategray4}      {rgb}{0.32,0.55,0.55}
\definecolor {aquamarine2}         {rgb}{0.46,0.93,0.78}
\definecolor {aquamarine4}         {rgb}{0.27,0.55,0.45}
\definecolor {darkseagreen1}       {rgb}{0.76,1.00,0.76}
\definecolor {darkseagreen2}       {rgb}{0.71,0.93,0.71}
\definecolor {darkseagreen3}       {rgb}{0.61,0.80,0.61}
\definecolor {darkseagreen4}       {rgb}{0.41,0.55,0.41}
\definecolor {seagreen1}           {rgb}{0.33,1.00,0.62}
\definecolor {seagreen2}           {rgb}{0.31,0.93,0.58}
\definecolor {seagreen3}           {rgb}{0.26,0.80,0.50}
\definecolor {palegreen1}          {rgb}{0.60,1.00,0.60}
\definecolor {palegreen2}          {rgb}{0.56,0.93,0.56}
\definecolor {palegreen3}          {rgb}{0.49,0.80,0.49}
\definecolor {palegreen4}          {rgb}{0.33,0.55,0.33}
\definecolor {springgreen2}        {rgb}{0.00,0.93,0.46}
\definecolor {springgreen3}        {rgb}{0.00,0.80,0.40}
\definecolor {springgreen4}        {rgb}{0.00,0.55,0.27}
\definecolor {green2}              {rgb}{0.00,0.93,0.00}
\definecolor {green3}              {rgb}{0.00,0.80,0.00}
\definecolor {green4}              {rgb}{0.00,0.55,0.00}
\definecolor {chartreuse2}         {rgb}{0.46,0.93,0.00}
\definecolor {chartreuse3}         {rgb}{0.40,0.80,0.00}
\definecolor {chartreuse4}         {rgb}{0.27,0.55,0.00}
\definecolor {olivedrab1}          {rgb}{0.75,1.00,0.24}
\definecolor {olivedrab2}          {rgb}{0.70,0.93,0.23}
\definecolor {olivedrab4}          {rgb}{0.41,0.55,0.13}
\definecolor {darkolivegreen1}     {rgb}{0.79,1.00,0.44}
\definecolor {darkolivegreen2}     {rgb}{0.74,0.93,0.41}
\definecolor {darkolivegreen3}     {rgb}{0.64,0.80,0.35}
\definecolor {darkolivegreen4}     {rgb}{0.43,0.55,0.24}
\definecolor {khaki1}              {rgb}{1.00,0.96,0.56}
\definecolor {khaki2}              {rgb}{0.93,0.90,0.52}
\definecolor {khaki3}              {rgb}{0.80,0.78,0.45}
\definecolor {khaki4}              {rgb}{0.55,0.53,0.31}
\definecolor {lightgoldenrod1}     {rgb}{1.00,0.93,0.55}
\definecolor {lightgoldenrod2}     {rgb}{0.93,0.86,0.51}
\definecolor {lightgoldenrod3}     {rgb}{0.80,0.75,0.44}
\definecolor {lightgoldenrod4}     {rgb}{0.55,0.51,0.30}
\definecolor {lightyellow2}        {rgb}{0.93,0.93,0.82}
\definecolor {lightyellow3}        {rgb}{0.80,0.80,0.71}
\definecolor {lightyellow4}        {rgb}{0.55,0.55,0.48}
\definecolor {yellow2}             {rgb}{0.93,0.93,0.00}
\definecolor {yellow3}             {rgb}{0.80,0.80,0.00}
\definecolor {yellow4}             {rgb}{0.55,0.55,0.00}
\definecolor {gold2}               {rgb}{0.93,0.79,0.00}
\definecolor {gold3}               {rgb}{0.80,0.68,0.00}
\definecolor {gold4}               {rgb}{0.55,0.46,0.00}
\definecolor {goldenrod1}          {rgb}{1.00,0.76,0.15}
\definecolor {goldenrod2}          {rgb}{0.93,0.71,0.13}
\definecolor {goldenrod3}          {rgb}{0.80,0.61,0.11}
\definecolor {goldenrod4}          {rgb}{0.55,0.41,0.08}
\definecolor {darkgoldenrod1}      {rgb}{1.00,0.73,0.06}
\definecolor {darkgoldenrod2}      {rgb}{0.93,0.68,0.05}
\definecolor {darkgoldenrod3}      {rgb}{0.80,0.58,0.05}
\definecolor {darkgoldenrod4}      {rgb}{0.55,0.40,0.03}
\definecolor {rosybrown1}          {rgb}{1.00,0.76,0.76}
\definecolor {rosybrown2}          {rgb}{0.93,0.71,0.71}
\definecolor {rosybrown3}          {rgb}{0.80,0.61,0.61}
\definecolor {rosybrown4}          {rgb}{0.55,0.41,0.41}
\definecolor {indianred1}          {rgb}{1.00,0.42,0.42}
\definecolor {indianred2}          {rgb}{0.93,0.39,0.39}
\definecolor {indianred3}          {rgb}{0.80,0.33,0.33}
\definecolor {indianred4}          {rgb}{0.55,0.23,0.23}
\definecolor {sienna1}             {rgb}{1.00,0.51,0.28}
\definecolor {sienna2}             {rgb}{0.93,0.47,0.26}
\definecolor {sienna3}             {rgb}{0.80,0.41,0.22}
\definecolor {sienna4}             {rgb}{0.55,0.28,0.15}
\definecolor {burlywood1}          {rgb}{1.00,0.83,0.61}
\definecolor {burlywood2}          {rgb}{0.93,0.77,0.57}
\definecolor {burlywood3}          {rgb}{0.80,0.67,0.49}
\definecolor {burlywood4}          {rgb}{0.55,0.45,0.33}
\definecolor {wheat1}              {rgb}{1.00,0.91,0.73}
\definecolor {wheat2}              {rgb}{0.93,0.85,0.68}
\definecolor {wheat3}              {rgb}{0.80,0.73,0.59}
\definecolor {wheat4}              {rgb}{0.55,0.49,0.40}
\definecolor {tan1}                {rgb}{1.00,0.65,0.31}
\definecolor {tan2}                {rgb}{0.93,0.60,0.29}
\definecolor {tan4}                {rgb}{0.55,0.35,0.17}
\definecolor {chocolate1}          {rgb}{1.00,0.50,0.14}
\definecolor {chocolate2}          {rgb}{0.93,0.46,0.13}
\definecolor {chocolate3}          {rgb}{0.80,0.40,0.11}
\definecolor {firebrick1}          {rgb}{1.00,0.19,0.19}
\definecolor {firebrick2}          {rgb}{0.93,0.17,0.17}
\definecolor {firebrick3}          {rgb}{0.80,0.15,0.15}
\definecolor {firebrick4}          {rgb}{0.55,0.10,0.10}
\definecolor {brown1}              {rgb}{1.00,0.25,0.25}
\definecolor {brown2}              {rgb}{0.93,0.23,0.23}
\definecolor {brown3}              {rgb}{0.80,0.20,0.20}
\definecolor {brown4}              {rgb}{0.55,0.14,0.14}
\definecolor {salmon1}             {rgb}{1.00,0.55,0.41}
\definecolor {salmon2}             {rgb}{0.93,0.51,0.38}
\definecolor {salmon3}             {rgb}{0.80,0.44,0.33}
\definecolor {salmon4}             {rgb}{0.55,0.30,0.22}
\definecolor {lightsalmon2}        {rgb}{0.93,0.58,0.45}
\definecolor {lightsalmon3}        {rgb}{0.80,0.51,0.38}
\definecolor {lightsalmon4}        {rgb}{0.55,0.34,0.26}
\definecolor {orange2}             {rgb}{0.93,0.60,0.00}
\definecolor {orange3}             {rgb}{0.80,0.52,0.00}
\definecolor {orange4}             {rgb}{0.55,0.35,0.00}
\definecolor {darkorange1}         {rgb}{1.00,0.50,0.00}
\definecolor {darkorange2}         {rgb}{0.93,0.46,0.00}
\definecolor {darkorange3}         {rgb}{0.80,0.40,0.00}
\definecolor {darkorange4}         {rgb}{0.55,0.27,0.00}
\definecolor {coral1}              {rgb}{1.00,0.45,0.34}
\definecolor {coral2}              {rgb}{0.93,0.42,0.31}
\definecolor {coral3}              {rgb}{0.80,0.36,0.27}
\definecolor {coral4}              {rgb}{0.55,0.24,0.18}
\definecolor {tomato2}             {rgb}{0.93,0.36,0.26}
\definecolor {tomato3}             {rgb}{0.80,0.31,0.22}
\definecolor {tomato4}             {rgb}{0.55,0.21,0.15}
\definecolor {orangered2}          {rgb}{0.93,0.25,0.00}
\definecolor {orangered3}          {rgb}{0.80,0.22,0.00}
\definecolor {orangered4}          {rgb}{0.55,0.15,0.00}
\definecolor {red2}                {rgb}{0.93,0.00,0.00}
\definecolor {red3}                {rgb}{0.80,0.00,0.00}
\definecolor {red4}                {rgb}{0.55,0.00,0.00}
\definecolor {deeppink2}           {rgb}{0.93,0.07,0.54}
\definecolor {deeppink3}           {rgb}{0.80,0.06,0.46}
\definecolor {deeppink4}           {rgb}{0.55,0.04,0.31}
\definecolor {hotpink1}            {rgb}{1.00,0.43,0.71}
\definecolor {hotpink2}            {rgb}{0.93,0.42,0.65}
\definecolor {hotpink3}            {rgb}{0.80,0.38,0.56}
\definecolor {hotpink4}            {rgb}{0.55,0.23,0.38}
\definecolor {pink1}               {rgb}{1.00,0.71,0.77}
\definecolor {pink2}               {rgb}{0.93,0.66,0.72}
\definecolor {pink3}               {rgb}{0.80,0.57,0.62}
\definecolor {pink4}               {rgb}{0.55,0.39,0.42}
\definecolor {lightpink1}          {rgb}{1.00,0.68,0.73}
\definecolor {lightpink2}          {rgb}{0.93,0.64,0.68}
\definecolor {lightpink3}          {rgb}{0.80,0.55,0.58}
\definecolor {lightpink4}          {rgb}{0.55,0.37,0.40}
\definecolor {palevioletred1}      {rgb}{1.00,0.51,0.67}
\definecolor {palevioletred2}      {rgb}{0.93,0.47,0.62}
\definecolor {palevioletred3}      {rgb}{0.80,0.41,0.54}
\definecolor {palevioletred4}      {rgb}{0.55,0.28,0.36}
\definecolor {maroon1}             {rgb}{1.00,0.20,0.70}
\definecolor {maroon2}             {rgb}{0.93,0.19,0.65}
\definecolor {maroon3}             {rgb}{0.80,0.16,0.56}
\definecolor {maroon4}             {rgb}{0.55,0.11,0.38}
\definecolor {violetred1}          {rgb}{1.00,0.24,0.59}
\definecolor {violetred2}          {rgb}{0.93,0.23,0.55}
\definecolor {violetred3}          {rgb}{0.80,0.20,0.47}
\definecolor {violetred4}          {rgb}{0.55,0.13,0.32}
\definecolor {magenta2}            {rgb}{0.93,0.00,0.93}
\definecolor {magenta3}            {rgb}{0.80,0.00,0.80}
\definecolor {magenta4}            {rgb}{0.55,0.00,0.55}
\definecolor {orchid1}             {rgb}{1.00,0.51,0.98}
\definecolor {orchid2}             {rgb}{0.93,0.48,0.91}
\definecolor {orchid3}             {rgb}{0.80,0.41,0.79}
\definecolor {orchid4}             {rgb}{0.55,0.28,0.54}
\definecolor {plum1}               {rgb}{1.00,0.73,1.00}
\definecolor {plum2}               {rgb}{0.93,0.68,0.93}
\definecolor {plum3}               {rgb}{0.80,0.59,0.80}
\definecolor {plum4}               {rgb}{0.55,0.40,0.55}
\definecolor {mediumorchid1}       {rgb}{0.88,0.40,1.00}
\definecolor {mediumorchid2}       {rgb}{0.82,0.37,0.93}
\definecolor {mediumorchid3}       {rgb}{0.71,0.32,0.80}
\definecolor {mediumorchid4}       {rgb}{0.48,0.22,0.55}
\definecolor {darkorchid1}         {rgb}{0.75,0.24,1.00}
\definecolor {darkorchid2}         {rgb}{0.70,0.23,0.93}
\definecolor {darkorchid3}         {rgb}{0.60,0.20,0.80}
\definecolor {darkorchid4}         {rgb}{0.41,0.13,0.55}
\definecolor {purple1}             {rgb}{0.61,0.19,1.00}
\definecolor {purple2}             {rgb}{0.57,0.17,0.93}
\definecolor {purple3}             {rgb}{0.49,0.15,0.80}
\definecolor {purple4}             {rgb}{0.33,0.10,0.55}
\definecolor {mediumpurple1}       {rgb}{0.67,0.51,1.00}
\definecolor {mediumpurple2}       {rgb}{0.62,0.47,0.93}
\definecolor {mediumpurple3}       {rgb}{0.54,0.41,0.80}
\definecolor {mediumpurple4}       {rgb}{0.36,0.28,0.55}
\definecolor {thistle1}            {rgb}{1.00,0.88,1.00}
\definecolor {thistle2}            {rgb}{0.93,0.82,0.93}
\definecolor {thistle3}            {rgb}{0.80,0.71,0.80}
\definecolor {thistle4}            {rgb}{0.55,0.48,0.55}
\definecolor {gray1}               {rgb}{0.01,0.01,0.01}
\definecolor {gray2}               {rgb}{0.02,0.02,0.02}
\definecolor {gray3}               {rgb}{0.03,0.03,0.03}
\definecolor {gray4}               {rgb}{0.04,0.04,0.04}
\definecolor {gray5}               {rgb}{0.05,0.05,0.05}
\definecolor {gray6}               {rgb}{0.06,0.06,0.06}
\definecolor {gray7}               {rgb}{0.07,0.07,0.07}
\definecolor {gray8}               {rgb}{0.08,0.08,0.08}
\definecolor {gray9}               {rgb}{0.09,0.09,0.09}
\definecolor {gray10}              {rgb}{0.10,0.10,0.10}
\definecolor {gray11}              {rgb}{0.11,0.11,0.11}
\definecolor {gray12}              {rgb}{0.12,0.12,0.12}
\definecolor {gray13}              {rgb}{0.13,0.13,0.13}
\definecolor {gray14}              {rgb}{0.14,0.14,0.14}
\definecolor {gray15}              {rgb}{0.15,0.15,0.15}
\definecolor {gray16}              {rgb}{0.16,0.16,0.16}
\definecolor {gray17}              {rgb}{0.17,0.17,0.17}
\definecolor {gray18}              {rgb}{0.18,0.18,0.18}
\definecolor {gray19}              {rgb}{0.19,0.19,0.19}
\definecolor {gray20}              {rgb}{0.20,0.20,0.20}
\definecolor {gray21}              {rgb}{0.21,0.21,0.21}
\definecolor {gray22}              {rgb}{0.22,0.22,0.22}
\definecolor {gray23}              {rgb}{0.23,0.23,0.23}
\definecolor {gray24}              {rgb}{0.24,0.24,0.24}
\definecolor {gray25}              {rgb}{0.25,0.25,0.25}
\definecolor {gray26}              {rgb}{0.26,0.26,0.26}
\definecolor {gray27}              {rgb}{0.27,0.27,0.27}
\definecolor {gray28}              {rgb}{0.28,0.28,0.28}
\definecolor {gray29}              {rgb}{0.29,0.29,0.29}
\definecolor {gray30}              {rgb}{0.30,0.30,0.30}
\definecolor {gray31}              {rgb}{0.31,0.31,0.31}
\definecolor {gray32}              {rgb}{0.32,0.32,0.32}
\definecolor {gray33}              {rgb}{0.33,0.33,0.33}
\definecolor {gray34}              {rgb}{0.34,0.34,0.34}
\definecolor {gray35}              {rgb}{0.35,0.35,0.35}
\definecolor {gray36}              {rgb}{0.36,0.36,0.36}
\definecolor {gray37}              {rgb}{0.37,0.37,0.37}
\definecolor {gray38}              {rgb}{0.38,0.38,0.38}
\definecolor {gray39}              {rgb}{0.39,0.39,0.39}
\definecolor {gray40}              {rgb}{0.40,0.40,0.40}
\definecolor {gray42}              {rgb}{0.42,0.42,0.42}
\definecolor {gray43}              {rgb}{0.43,0.43,0.43}
\definecolor {gray44}              {rgb}{0.44,0.44,0.44}
\definecolor {gray45}              {rgb}{0.45,0.45,0.45}
\definecolor {gray46}              {rgb}{0.46,0.46,0.46}
\definecolor {gray47}              {rgb}{0.47,0.47,0.47}
\definecolor {gray48}              {rgb}{0.48,0.48,0.48}
\definecolor {gray49}              {rgb}{0.49,0.49,0.49}
\definecolor {gray50}              {rgb}{0.50,0.50,0.50}
\definecolor {gray51}              {rgb}{0.51,0.51,0.51}
\definecolor {gray52}              {rgb}{0.52,0.52,0.52}
\definecolor {gray53}              {rgb}{0.53,0.53,0.53}
\definecolor {gray54}              {rgb}{0.54,0.54,0.54}
\definecolor {gray55}              {rgb}{0.55,0.55,0.55}
\definecolor {gray56}              {rgb}{0.56,0.56,0.56}
\definecolor {gray57}              {rgb}{0.57,0.57,0.57}
\definecolor {gray58}              {rgb}{0.58,0.58,0.58}
\definecolor {gray59}              {rgb}{0.59,0.59,0.59}
\definecolor {gray60}              {rgb}{0.60,0.60,0.60}
\definecolor {gray61}              {rgb}{0.61,0.61,0.61}
\definecolor {gray62}              {rgb}{0.62,0.62,0.62}
\definecolor {gray63}              {rgb}{0.63,0.63,0.63}
\definecolor {gray64}              {rgb}{0.64,0.64,0.64}
\definecolor {gray65}              {rgb}{0.65,0.65,0.65}
\definecolor {gray66}              {rgb}{0.66,0.66,0.66}
\definecolor {gray67}              {rgb}{0.67,0.67,0.67}
\definecolor {gray68}              {rgb}{0.68,0.68,0.68}
\definecolor {gray69}              {rgb}{0.69,0.69,0.69}
\definecolor {gray70}              {rgb}{0.70,0.70,0.70}
\definecolor {gray71}              {rgb}{0.71,0.71,0.71}
\definecolor {gray72}              {rgb}{0.72,0.72,0.72}
\definecolor {gray73}              {rgb}{0.73,0.73,0.73}
\definecolor {gray74}              {rgb}{0.74,0.74,0.74}
\definecolor {gray75}              {rgb}{0.75,0.75,0.75}
\definecolor {gray76}              {rgb}{0.76,0.76,0.76}
\definecolor {gray77}              {rgb}{0.77,0.77,0.77}
\definecolor {gray78}              {rgb}{0.78,0.78,0.78}
\definecolor {gray79}              {rgb}{0.79,0.79,0.79}
\definecolor {gray80}              {rgb}{0.80,0.80,0.80}
\definecolor {gray81}              {rgb}{0.81,0.81,0.81}
\definecolor {gray82}              {rgb}{0.82,0.82,0.82}
\definecolor {gray83}              {rgb}{0.83,0.83,0.83}
\definecolor {gray84}              {rgb}{0.84,0.84,0.84}
\definecolor {gray85}              {rgb}{0.85,0.85,0.85}
\definecolor {gray86}              {rgb}{0.86,0.86,0.86}
\definecolor {gray87}              {rgb}{0.87,0.87,0.87}
\definecolor {gray88}              {rgb}{0.88,0.88,0.88}
\definecolor {gray89}              {rgb}{0.89,0.89,0.89}
\definecolor {gray90}              {rgb}{0.90,0.90,0.90}
\definecolor {gray91}              {rgb}{0.91,0.91,0.91}
\definecolor {gray92}              {rgb}{0.92,0.92,0.92}
\definecolor {gray93}              {rgb}{0.93,0.93,0.93}
\definecolor {gray94}              {rgb}{0.94,0.94,0.94}
\definecolor {gray95}              {rgb}{0.95,0.95,0.95}
\definecolor {gray97}              {rgb}{0.97,0.97,0.97}
\definecolor {gray98}              {rgb}{0.98,0.98,0.98}
\definecolor {gray99}              {rgb}{0.99,0.99,0.99}
\definecolor {darkgrey}            {rgb}{0.66,0.66,0.66}

%% file: rs_macros_mc.tex
\newcommand{\false}{\bot}

%% file: rs_macros_optsmt.tex
\newcommand{\omt}{\ensuremath{\text{OMT}}\xspace}

\newcommand{\omlaratplus}{\ensuremath{\text{OMT}(\larat\cup\T)}\xspace}

\newcommand{\cost}{\ensuremath{{cost}}\xspace}

\newcommand{\C}{\ensuremath{\mathcal{C}}\xspace}

\newcommand\mysout{\bgroup \markoverwith{{-}}\ULon}
\newcommand\nosout{\bgroup \markoverwith{{ }}\ULon}
\definecolor{mygray}{rgb}{0.90,0.90,0.90}
\definecolor{mywhite}{rgb}{1.00,1.00,1.00}

\newcommand{\optimathsat}{\textsc{OptiMathSAT}\xspace}

\newcommand{\symba}{\textsc{Symba}\xspace}
\newcommand{\nuz}{\textsc{$\nu Z$}\xspace}
\newcommand{\bclt}{\textsc{bclt}\xspace}

%% file: rs_macros_sat.tex
\newcommand{\bcp}{\ensuremath{\sf{boolean\_constraint\_propagation}}\xspace}

\newcommand{\unsatres}{\textsc{unsat}\xspace}

%% file: rs_macros_smt.tex
\newcommand{\vi}{\ensuremath{\varphi}\xspace}

\newcommand{\T}{\ensuremath{\mathcal{T}}\xspace}

\newcommand{\smt}{SMT\xspace}

\newcommand{\larat}{\ensuremath{\mathcal{LA}(\mathbb{Q})}\xspace}

\renewcommand{\larat}{\ensuremath{\mathcal{LRA}}\xspace}

\newcommand{\TsolverGen}[1]{\ensuremath{{#1}\textit{-solver}}\xspace}
\newcommand{\TsolversGen}[1]{\ensuremath{{#1}\textit{-solvers}}\xspace}
\newcommand{\Tsolver}{\TsolverGen{\T}}
\newcommand{\Tsolvers}{\TsolversGen{\T}}

\newcommand{\laratsolver}{\larat-\ensuremath{\mathsf{Solver}}}

\newcommand{\yices}{\textsc{Yices}\xspace}

\newcommand{\zthree}{\textsc{Z3}\xspace}

%% file: rs_macros_specific.tex
\newcommand{\omtpb}{\ensuremath{\text{OMT+PB}}\xspace}

%% file: pt_macros.tex
\newcommand{\PTNOTE}[1]{{\textcolor{darkviolet}{\textbf{PT: }
      {\footnotesize #1}}}}
\newcommand{\minmax}{\textsc{min-max}\xspace}
\newcommand{\maxmin}{\textsc{max-min}\xspace}
\renewcommand{\minmax}{min-max\xspace}
\renewcommand{\maxmin}{max-min\xspace}
\newcommand{\maxsmt}{\textsc{MaxSMT}\xspace}
\renewcommand{\cost}{\ensuremath{{obj}}\xspace}

\renewcommand{\nuz}{\textsc{$\nu Z$}\xspace}
\newcommand{\nuZ}{\textsc{$\nu Z$}\xspace}
\newcommand{\vZ}{\textsc{$\nu Z$}\xspace}
\newcommand{\maxino}{\textsc{Maxino}\xspace}

\newcommand{\maxsat}{\textsc{MaxSAT}\xspace}

%% file: abstract.tex
\noi
Optimization Modulo Theories (\omt) is an extension of \smt 
which allows 
for finding models that optimize given
objectives.
%
% \ignore{
% \omt has been extended to be incremental and to handle multiple
% objective functions either independently or with their linear,
% lexicographic, Pareto, \minmax/\maxmin combinations.
% }
%
\ignore{
\omt applications can be found not only in the domain of Formal
Verification and related domains, 
Automated 
Reasoning and Planning with Resources, but also in other disciplines
like Machine Learning and
Requirement Engineering. 
}
(Partial weighted) \maxsmt \\--or equivalently \omt with
Pseudo-Boolean objective functions, \omtpb{}-- 
is a very-relevant strict subcase of \omt.  
%
% \omt with
% Pseudo-Boolean objective functions, \omtpb{} 
%  --or equivalently 
% (partial weighted) \maxsmt-- 
% is a very-relevant subcase of \omt.  
%
We classify existing approaches for \maxsmt or \omtpb in 
two groups\ignoreinshort{, each
with specific limitations}:
{\em \maxsat-based} approaches 
 exploit the 
efficiency of state-of-the-art \maxsat solvers, 
but they are specific-purpose and not always applicable;
{\em \omt-based} approaches are general-purpose, but they suffer from
 intrinsic inefficiencies on  \maxsmt/\omtpb problems.

\ignore{In this paper }
We identify a major source of such inefficiencies, and we address it by
enhancing \omt by means of bidirectional 
sorting networks. 
{We implemented
this idea on top of the \optimathsat \omt solver.}
We run an extensive empirical evaluation on a variety of 
problems\ignoreinshort{ coming from Machine Learning and
Requirement Engineering}, 
comparing \maxsat-based and \omt-based
 techniques, with and without sorting networks,
 implemented on top of \optimathsat{} and \nuz. 
The results support the effectiveness of this idea, and provide
interesting insights about the different approaches. 

%  and evaluated them
% empirically on problems coming from Machine Learning and
% Requirement Engineering.
%
% The empirical results 

%% file: intro.tex
Satisfiability Modulo Theories (\smt) is the problem of deciding the
satisfiability of first-order formulas with respect to 
first-order theories \cite{sebastiani07,BSST09HBSAT} 
(e.g., the theory of linear arithmetic over the rationals, \larat). In the last
decade, \smt solvers --powered by very efficient
Conflict-Driven-Clause-Learning (CDCL) engines for Boolean
Satisfiability \cite{MSLM09HBSAT} combined with a collection of
\T-Solvers, each one handling a different theory \T-- have risen to be
a pervasive and indispensable tool for dealing with many problems of
industrial interest, e.g.  formal verification of hardware and
software systems, resource planning, temporal reasoning and scheduling
of real-time embedded systems.

Optimization Modulo Theories (\omt) is an extension of \smt, which
allows for finding models that make a given objective optimum through
a combination of \smt and optimization procedures 
\cite{nieuwenhuis_sat06,cimattifgss10,cgss_sat13_maxsmt,st-ijcar12,st_tocl14,larrazorr14,li_popl14,bjorner_scss14,bjorner_tacas15,st_tacas15,st_cav15}.
%
%\marg{Eventualmente tagliare}
Latest advancements in \omt have further broadened its horizon by
making it incremental \cite{bjorner_scss14,st_tacas15} and by
supporting objectives defined in other theories than linear arithmetic
(e.g. Bit-Vectors)
\cite{bjorner_scss14,bjorner_tacas15,NadelR16}. Moreover, \omt has
been extended with the capability of handling multiple objectives at
the same time either independently (aka boxed optimization) or through their linear,
\minmax/\maxmin, lexicographic or Pareto combination
\cite{bjorner_scss14,bjorner_tacas15,st_tacas15}.

We focus on an important strict sub-case of \omt,  (partial
weighted)\footnote{Hereafter, when speaking of \maxsat or \maxsmt, we
keep ``partial weighted'' implicit.}  \maxsmt{}
 --or equivalently \omt with Pseudo-Boolean (PB)
objective functions \cite{RM09HBSAT}, \omtpb--
 which is the
problem of finding a model for an input formula which both satisfies
all \textit{hard} clauses and maximizes the cumulative weight of all
\textit{soft} clauses satisfied by the model 
\cite{nieuwenhuis_sat06,cimattifgss10,cgss_sat13_maxsmt}.
We identify
two main approaches which have been adopted in the literature (see
related work).
One specific-purpose approach, which we call \maxsat{}{\em -based},
 is to embed some \maxsat engine within the \smt solver
itself, and use it in combination with dedicated \Tsolvers 
\cite{AnsoteguiBPSV11,bjorner_scss14,bjorner_tacas15}
or with \smt solvers used as blackboxes
\cite{cgss_sat13_maxsmt}.
One general-purpose approach, which we call {\em \omt-based}, is to encode \maxsmt{}/\omtpb into general 
\omt with linear-real-arithmetic cost
functions \cite{st_tocl14}.

We compare the two approaches and notice the following facts. 

The \maxsat-based approach can 
% be very efficient --as
%  we will show empirically in this paper-- by exploiting 
exploit the efficiency of state-of-the-art
  \maxsat procedures and solvers. Unfortunately it suffers from some 
 limitations that make it impractical or inapplicable   in some
 cases.
First, to the best of our knowledge, available \maxsat engines
deal with integer weights only;  % and, unlike \omt, they might suffer
some applications, e.g., {\em (Machine) Learning Modulo Theories, LMT}
\cite{teso_aij15} --a hybrid Machine Learning approach in
which \omt is used as an oracle for Support Vector Machines
\cite{teso_aij15}-- may require the weight of 
soft constraints to be high-precision rational values.~%
\footnote{For example, %the weight 
$\frac{1799972218749879}{2251799813685248}$
\ignoreinshort{, which expands to $0.79934824037669782725856748584192...$, }
is a sample weight value from problems in \cite{teso_aij15}.}
% obtained from a previous iteration of the \omt search. 
(In this context, it is preferable not to round the weights associated
with {soft}-clauses since it affects the accuracy of the
Machine Learning approach;~%
also multiplying all rational
coefficients for their lowest common 
multiple of the
 denominators is not practical,
because such values tend to become huge.)
% , 
% and most \maxsat solvers can afford only limited-size 
% integer weights (e.g. 32-bit integers 
% for most solvers in the \maxsat competition.)
%
Second, a \maxsat engine cannot be directly used
when dealing with an \omt problem with multiple-independent objectives
that need to be optimized at the same time \cite{li_popl14},%
\footnote{One could run a \maxsat-based{}
search separately on each objective, but doing this he/she would loose the benefits 
of boxed optimization, see \cite{li_popl14,bjorner_scss14,st_tacas15}.} 
or when the objective 
function is given by combinations of PB and
arithmetic terms --like, e.g., for Linear Generalized Disjunctive
Programming problems \cite{sawayag05,st_tocl14} or LMT problems \cite{teso_aij15}.

The \omt-based approach does not suffer from the above
limitations, because it exploits the infinite-precision
linear-arithmetic package on the rationals of OMT solvers,
and it treats PB functions as any other arithmetic
functions.
%
% To cope with these facts, 
% an alternative approach for dealing with \maxsmt is to encode it as a
% PB objective in Optimization Modulo Theories
% \cite{st_tocl14}.
 % which, for the above reasons,
 % is one of the  approaches adopted in \optimathsat \cite{optimathsat-url}.
Nevertheless %here we notice that
%compared with the use of a dedicated \maxsat engine, 
this approach may result in low performances
when dealing with \maxsmt{}/\omtpb problems.
\ignore{
 problems in which %significant amounts of
\textit{soft}-clauses are assigned the same weight. 
}

We analyze the latter fact and identify a major source of inefficiency
by noticing that the presence of same-weight soft clauses
entails the existence of symmetries in the solution space
that may lead to a combinatorial explosion of the partial truth
assignments generated by the CDCL engine during the optimization
search.
To cope with this fact, we introduce and describe
 a solution based on (bidirectional) sorting networks \cite{Sinz05,Asin2011,parametricCardinalityConstraint}. We
implemented this idea within the \optimathsat \omt solver \cite{st_cav15}.

We run an empirical evaluation on a large amount
of problems 
\ignoreinshort{coming from Machine Learning and
Requirement Engineering, }
comparing 
\maxsat-based and \omt-based
 techniques, with and without sorting networks,
 implemented on top of \optimathsat{}  \cite{st_cav15} 
and \nuz  %(aka \textsc{Z3Opt}) 
\cite{bjorner_tacas15}. 
The results are summarized as follows.
\begin{aenumerate}
\item Comparing  \maxsat-based wrt. \omt-based approaches  on 
 problems where the former are applicable, it turns out that the
 former provide much better performances, in particular when adopting
 the maximum-resolution 
\cite{narodytskab14,bjorner_scss14} \maxsat{} engine. 
\item Evaluating the
benefits of bidirectional 
sorting-network encodings, it turns out 
that they improve significantly the performances 
of \omt-based approaches, 
%most often when adopting the cardinality-network encodings of \cite{Asin2011}, 
and
often also of \maxsat-based ones.
\item
Comparing \nuz and \optimathsat, it turns out that the former 
performed better on \maxsat{}-based approaches, whilst 
the latter performed seomtimes equivalently and sometimes
 significantly better on \omt-based ones, in particular when enhanced
 by the sorting-network encoding. 
\end{aenumerate}

\paragraph{Related Work.} %
\input{related}

\paragraph{Content.} 
The paper is structured as
follows. 
\sref{sec:background} briefly reviews
the background; %and the state of the art 
%% is briefly reviewed
%% in \sref{sec:background}, whereas 
\sref{sec:problem}
describes the source of inefficiency arising when %partial weighted
\maxsmt is encoded in \omt as in \cite{st_tocl14};
\sref{sec:combining} illustrates a possible solution based on
bidirectional sorting
networks; in \sref{sec:expeval} we provide empirical evidence
of the benefits of this approach on two applications of \omt
interest. \sref{sec:concl} provides some conclusions with some
considerations on the future work.

\paragraph{Note to reviewers.} 
This paper extends a previous paper  \cite{st_smt16} which 
was presented
at SMT'16 workshop and published in its informal
proceedings. Nevertheless, given the informal and non-archival
nature of those proceedings, we believe that 
the current submission 
complies with the TACAS policies on the originality of the contribution.

%% file: related.tex
The idea of MaxSMT and\ignoreinshort{, more generally,}
of optimization in SMT was
first introduced by Nieuwenhuis \& Oliveras \cite{nieuwenhuis_sat06},
who presented a general logical framework of ``SMT with progressively
stronger theories'' (e.g., where the theory is progressively
strengthened by every new approximation of the minimum cost), and
presented implementations for MaxSMT based on this framework.
Cimatti et al. \cite{cimattifgss10} introduced the notion of ``Theory
of Costs'' \calc to handle Pseudo-Boolean (PB) cost functions and
constraints by an ad-hoc \ignoreinshort{and independent} ``\C-solver'' in the standard
lazy SMT schema, and implemented a variant of MathSAT tool able to
handle SMT with PB constraints and to minimize PB cost functions.
\ignoreinshort{
Ans{\'o}tegui et al. \cite{AnsoteguiBPSV11} described the evaluation of
an implementation of a MaxSMT procedure based on \yices, although
this implementation is not publicly available.
}
Cimatti et al. \cite{cgss_sat13_maxsmt} presented a ``modular''
approach for MaxSMT, combining a lazy \smt{} solver with a MaxSAT
solver, \ignoreinshort{which can be used as black-boxes,}
 where the \smt solver is used
as an oracle generating \T-lemmas that are then learned by the
\textsc{MaxSAT} solver so as to progressively narrow the search space
toward the optimal solution.

Sebastiani and Tomasi \cite{st-ijcar12,st_tocl14} introduced 
a wider notion of optimization in SMT, namely {\em
Optimization Modulo Theories (OMT) with \larat cost functions}, \omlaratplus, which allows for finding models minimizing some \larat cost term
--\T being some (possibly empty) stably-infinite theory s.t. \T and
\larat are signature-disjoint-- and presented novel 
 \omlaratplus{} tools which combine standard SMT
with LP minimization techniques.  
%\footnote{
(\T can also be a combination of Theories
$\bigcup_i\T_i$.)
%}
\ignoreinshort{
% Generally speaking, these approaches can only deal with a class of problems that is stricter than the one targeted by \omt. 
Importantly, both \maxsmt and \smt with PB objectives can be
encoded into \omlaratplus, 
whereas the contrary is not possible \cite{st-ijcar12,st_tocl14}.
}
Eventually, \omlaratplus{} has been extended so that to handle 
costs on the integers,
incremental OMT, 
multi-objective, 
%min/max 
 and lexicographic OMT and Pareto-optimality 
 \cite{li_popl14,larrazorr14,bjorner_scss14,st_tacas15,bjorner_tacas15,st_cav15}. 
To the best of our knowledge only four OMT solvers are currently implemented:
\bclt \cite{larrazorr14}, \nuz (aka \textsc{Z3Opt})
 \cite{bjorner_scss14,bjorner_tacas15}, \optimathsat
 \cite{st_tacas15,st_cav15},  and 
 \symba \cite{li_popl14}.
Remarkably, \bclt{}, \nuz and \optimathsat{} currently implement also
specialized procedures for MaxSMT, leveraging to SMT level
state-of-the-art MaxSAT procedures; in addition, \nuz features a
Pseudo-Boolean \Tsolver which can generate sorting circuits on demand
for Pseudo-Boolean inequalities featuring sums with small coefficients
when a Pseudo-Boolean inequality is used some times for unit
propagation/conflicts \cite{bjorner_tacas15,bjorner_private16}.

%% file: background.tex
We assume the reader is familiar with the main theoretical and
algorithmic concepts in SAT and SMT solving (see
\cite{MSLM09HBSAT,BSST09HBSAT}).
Optimization Modulo Theories (\omt) is an extension of \smt which addresses
the problem of finding a model for an input formula $\varphi$ which is
optimal wrt. some objective function $\cost$
\cite{nieuwenhuis_sat06,st-ijcar12}.
The basic minimization scheme implemented in state-of-the-art \omt
solvers, known as \textit{linear-search} scheme
\cite{nieuwenhuis_sat06,st-ijcar12}, requires solving an \smt problem with a
solution space that is progressively tightened by means of unit linear
constraints in the form $\neg(ub_i \leq \cost)$, where $ub_i$ is the value of
$\cost$ that corresponds to the optimum model of the most-recently
found truth assignment $\mu_i$ s.t. $\mu_i \models \varphi$. The
$ub_i$ value is computed by means of a specialized optimization
procedure embedded within the $\Tsolver$ which, taken as
input a pair $\langle \mu, \cost \rangle$, returns the optimal value
$ub$ of $\cost$ for such $\mu$.  The \omt search terminates when such
procedure finds that $\cost$ is unbounded or when the \smt
search is $\unsatres$, in which case the latest value of $\cost$ (if
any) and its associated model $M_i$ is returned as optimal solution
value. (Alternatively, binary-search schemes can also be used
\cite{st-ijcar12,st_tocl14}.)

An important subcase of \omt is that of \maxsmt, which is a pair
$\langle\varphi_h, \varphi_s\rangle$, where $\varphi_h$ denotes the
set of ``hard'' \T-clauses, $\varphi_s$ is a set of positive-weighted
``soft'' \T-clauses, and the goal is to find the maximum-weight set of
\T-clauses $\psi_s$, $\psi_s\subseteq\varphi_s$, s.t.  $\varphi_h \cup
\psi_s$ is \T-satisfiable \cite{nieuwenhuis_sat06,cimattifgss10,AnsoteguiBPSV11,cgss_sat13_maxsmt}.
As described in \cite{st_tocl14}, \maxsmt
$\langle\varphi_h,\varphi_s\rangle$ can be encoded into a general \omt problem 
with a Pseudo-Boolean objective: 
first introduce a fresh Boolean 
variable $A_i$ for each soft-constraint $C_i\in\varphi_s$ as follows
\begin{eqnarray}\label{eq:maxsmt}
\textstyle
\varphi^* \defas \varphi_h \cup \bigcup_{C_i\in\varphi_s}\{(A_i\vee C_i)\}; \:\: \cost \defas \sum_{C_i\in\varphi_s}w_i A_i
\end{eqnarray}
\noi
and then encode the problem into \omt as a pair $\langle \varphi,
\cost\rangle$ where $\varphi$ is
defined as
\begin{eqnarray}
\label{eq:maxsmt2}
\varphi &\defas& 
\textstyle
\varphi^* \wedge 
\bigwedge_{i} ((\neg A_i \vee (x_i=w_i)) \wedge (A_i \vee (x_i=0)))
\wedge \\
\label{eq:extraconstraints}
&& \textstyle
\bigwedge_{i} ((0 \le x_i) \wedge (x_i \le w_i)) \wedge\\
\label{eq:maxsmt3}
&&\textstyle
(\cost  = 
\sum_i x_i),\ \ \ \ \ x_i,\ \cost\ fresh.
% \cost & \defas & \textstyle
% \sum_i x_i,\ x_i\ fresh
\end{eqnarray}
\noi Notice that, although redundant from a logical perspective, the
constraints in \eqref{eq:extraconstraints} serve the important purpose
of allowing early-pruning calls to the \laratsolver{} (see
\cite{BSST09HBSAT}) to detect a possible \larat inconsistency among
the current partial truth assignment over variables $A_i$ and linear
cuts in the form $\neg(ub \leq \cost)$ that are pushed on the formula stack
by the \omt solver during the minimization of $\cost$.  To this
extent, the presence of such constraints improves performance significantly.

%% file: problem.tex
Consider first the case of a \maxsmt-derived \omt problem as in
 \eqref{eq:maxsmt} s.t.  all weights are identical, that is: let
$\langle \varphi, \cost \rangle$ be an \omt problem, where $\cost =
\sum_{i=1}^{n} w \cdot A_i$, where the $A_i$s are Boolean variables, and let
$\mu$ be a satisfiable truth assignment found by the \omt solver
during the minimization of $\cost$.
Given $A_{T}=\{A_i | \mu\models A_i\}$ and $k = |A_{T}|$, then 
the upper bound value of $\cost$ in $\mu$ is $ub = w\cdot k$.
As described in \cite{st-ijcar12,st_tocl14}, the \omt solver adds a
unit clause in the form $\neg(ub \leq \cost)$ in order to (1) remove
the current truth assignment $\mu$ from the feasible search space and
(2) seek for another $\mu'$ which improves the current upper-bound
value $ub$.
Importantly, the unit clause $\neg(ub \leq \cost)$ does not only prune
the current truth assignment $\mu$ from the feasible search space, but
it also makes inconsistent any other  (partial) truth
assignment $\mu'$ which sets exactly $k$ (or more) $A_i$ variables to
True.
Thus, each new unit clause in this form prunes  $\gamma =
{{n}\choose{k}}$ truth assignments from the search space, where
$\gamma$ is the number of of possible
permutations of $\mu$ over the variables $A_i$.
A dual case occurs when some lower-bound unit clause $\neg(\cost \leq
lb)$ is learned (e.g., in a binary-search step, see \cite{st-ijcar12}).

Unfortunately, the inconsistency of a truth assignment $\mu'$ which sets
exactly $k$ variables to True wrt. a unit clause $\neg(ub \leq
\cost)$, where $ub = w \cdot k$, cannot be determined by simple
Boolean Constraint Propagation (BCP).
In fact, $\neg(ub \leq \cost)$ being a \larat term, the CDCL engine is
totally oblivious to this inconsistency until when the \Tsolver{} for
linear rational arithmetic (\laratsolver{}) is invoked, and a conflict clause is generated.
Therefore, since the \laratsolver{}  is much more 
resource-demanding than BCP and it is invoked less often, it is
clear that the performance of an \omt solver can be negatively
affected when dealing with this kind of objectives.

%\newpage
\begin{figure}[tb]
	\begin{center}
    \def\svgwidth{0.8\columnwidth}
    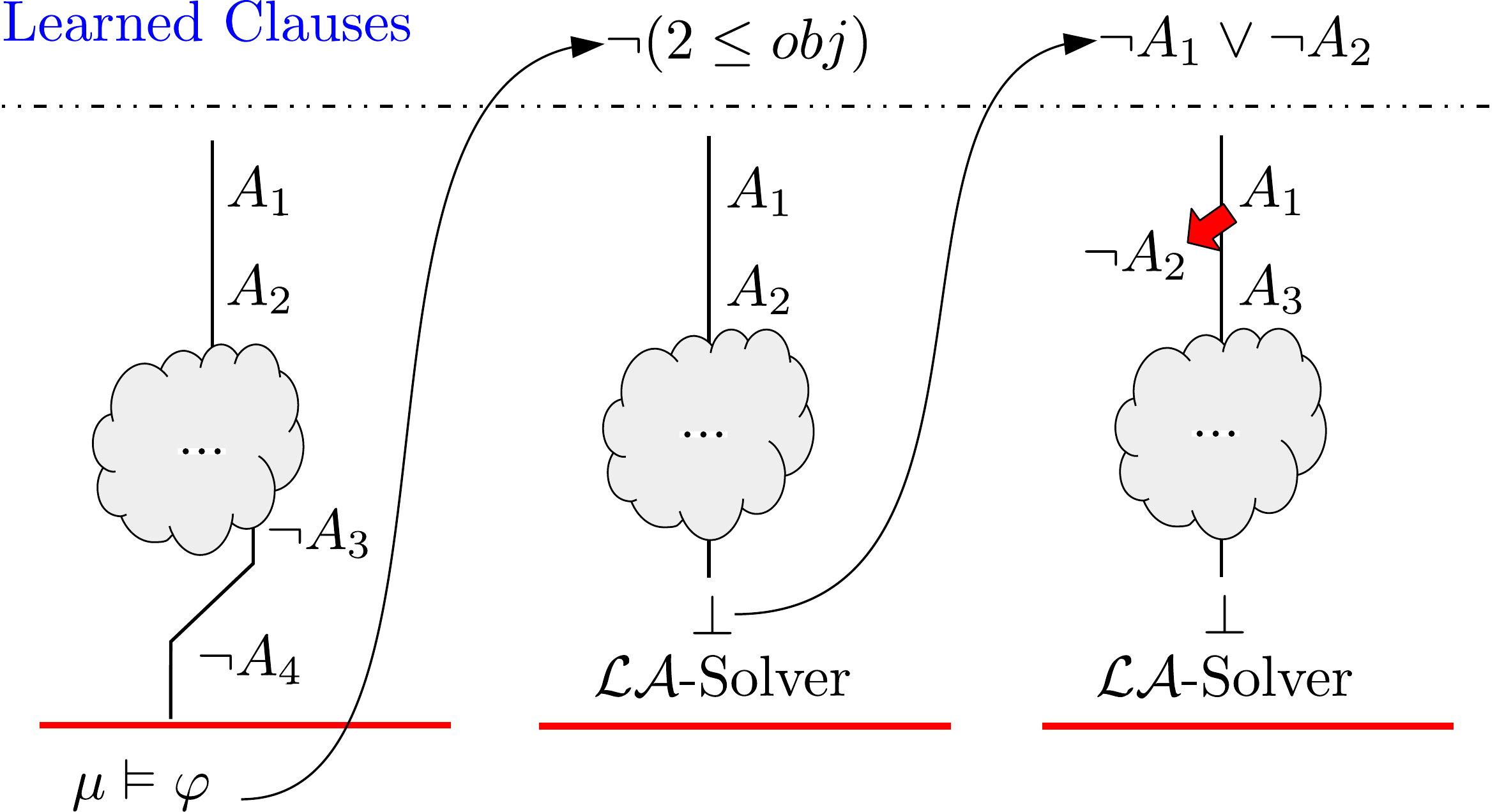
	\end{center}
\caption{\label{fig:prob3}
	\textmd{A simple example of \omt search.
	}
}
\end{figure}

\begin{example}
\label{ex:nocardinalitynetwork}
  Figure \ref{fig:prob3} shows a toy example of \omt search execution
  over the pair $\langle \varphi, \cost \rangle$, where $\vi$ is some
  SMT formula and $\cost \defas
  \sum_{i=1}^{4} A_i$ (i.e., $w_i=1$ for every $i$). We assume the problem has been encoded as in 
\eqref{eq:maxsmt2}-\eqref{eq:maxsmt3}, so that 
the truth assignment
$\mu_0\defas\cup_{i=1}^4\set{(0\le x_i),(x_i\le 1)}\cup\set{(\cost=\sum_{i=1}^4x_i)}$
is immediately generated by BCP, and is part of all truth
assignments generated in the search.
  In the first branch  (left) a truth
  assignment 
$\mu\defas\mu_0\cup\set{A_1,(x_1=1),A_2,(x_2=1),\neg A_3,(x_3=0),\neg A_4,(x_4=0)}$  is found s.t. $\cost = 2$, resulting from the decisions
  $A_1$, $A_2$, $\neg A_3$ and $\neg A_4$. Then the
  unit clause $\neg(2\leq\cost)$ is learned and the Boolean search is
  restarted in order to find an improved solution. 
  In the second branch   (center) $A_1$ and $A_2$ are decided, forcing
  by BCP   the assignment 
$\mu'\defas\mu_0\cup\set{\neg(2\leq\cost),A_1,(x_1=1),A_2,(x_2=1)}$
which is \larat-inconsistent.
However, it takes a (possibly-expensive) intermediate  call to the
\laratsolver{} 
to reveal such an inconsistency.~\footnote{The fact that such call 
is actually performed depends on the early-pruning strategy
implemented in the OMT solver; 
%\cite{BSST09HBSAT}.  
alternatively, a possibly-expensive \T-propagation step 
on the previous \laratsolver{} call has a similar effect. 
(See e.g. \cite{sebastiani07,BSST09HBSAT}.)} 
If so, a new conflict clause $\neg A_1
  \vee \neg A_2$ is learned, 
 forcing the solver to back-jump and toggle the value of $A_2$
(right). The search
  continues with the new decision $A_3$, which is again \larat{}
  inconsistent, causing a new conflict clause as before, and so
  on. In this way, the solver might uselessly enumerate and check all the 
up-to ${{4}\choose{2}}$ assignments that assign two $A_i$'s to true
and are consistent with \vi, 
even though they are intrinsically incompatible with  $\neg(2\leq\cost)$. 
\hfill $\diamond$ 
\end{example}

The performance issue identified with the previous case example can be
generalized to any objective $\cost$ 
in which groups of $A_{i}$'s share the same weights:
\begin{eqnarray}
\label{eq:gen1}
&&\cost = \tau_1 + ... + \tau_m, \\
\label{eq:gen2}
&&\textstyle
%\forall_j \in [1, m].:
\bigwedge_{j=1}^m\ (\ (\tau_j = w_j \cdot \sum_{i=1}^{k_j}A_{ji}) \:\:
\wedge \:\:(0 \leq \tau_j) \wedge (\tau_j \leq w_j \cdot k_j)\ ),
\end{eqnarray}
% \[
% \cost = \tau_1 + ... + \tau_m, \\
% \]
\noi
where the logically-redundant constraints $(0 \leq \tau_j) \wedge (\tau_j \leq w_j \cdot k_j)$ are added for 
the same reason as with \eqref{eq:extraconstraints}.

% \[
% \forall_j \in [1, m].\:\tau_j = w_j \cdot \sum_{i=0}^{i=k_j}A_i \:\:
% \wedge \:\:0 \leq \tau_j \leq w_j \cdot k_j
% \]

%% file: png/1.pdf_tex
%% Creator: Inkscape inkscape 0.91, www.inkscape.org
%% PDF/EPS/PS + LaTeX output extension by Johan Engelen, 2010
%% Accompanies image file '1.pdf' (pdf, eps, ps)
%%
%% To include the image in your LaTeX document, write
%%   \input{<filename>.pdf_tex}
%%  instead of
%%   \includegraphics{<filename>.pdf}
%% To scale the image, write
%%   \def\svgwidth{<desired width>}
%%   \input{<filename>.pdf_tex}
%%  instead of
%%   \includegraphics[width=<desired width>]{<filename>.pdf}
%%
%% Images with a different path to the parent latex file can
%% be accessed with the `import' package (which may need to be
%% installed) using
%%   \usepackage{import}
%% in the preamble, and then including the image with
%%   \import{<path to file>}{<filename>.pdf_tex}
%% Alternatively, one can specify
%%   \graphicspath{{<path to file>/}}
%% 
%% For more information, please see info/svg-inkscape on CTAN:
%%   http://tug.ctan.org/tex-archive/info/svg-inkscape
%%
\begingroup%
  \makeatletter%
  \providecommand\color[2][]{%
    \errmessage{(Inkscape) Color is used for the text in Inkscape, but the package 'color.sty' is not loaded}%
    \renewcommand\color[2][]{}%
  }%
  \providecommand\transparent[1]{%
    \errmessage{(Inkscape) Transparency is used (non-zero) for the text in Inkscape, but the package 'transparent.sty' is not loaded}%
    \renewcommand\transparent[1]{}%
  }%
  \providecommand\rotatebox[2]{#2}%
  \ifx\svgwidth\undefined%
    \setlength{\unitlength}{674.78743453bp}%
    \ifx\svgscale\undefined%
      \relax%
    \else%
      \setlength{\unitlength}{\unitlength * \real{\svgscale}}%
    \fi%
  \else%
    \setlength{\unitlength}{\svgwidth}%
  \fi%
  \global\let\svgwidth\undefined%
  \global\let\svgscale\undefined%
  \makeatother%
  \begin{picture}(1,0.54274306)%
    \put(0,0){\includegraphics[width=\unitlength,page=1]{png/1.pdf}}%
  \end{picture}%
\endgroup%

%% file: combining.tex
\newcommand{\ins}{\ensuremath{\underline{A}}}
\newcommand{\outs}{\ensuremath{\underline{B}}}
\newcommand{\sn}{\ensuremath{{\sf SN}[\ins,\outs]}}
\renewcommand{\bcp}[3]{\ensuremath{\tuple{#2,#1}\vdash_{{\sf bcp}}#3}}

\newcommand{\bcpcompliant}{BCP-compliant}
\newcommand{\bidirectional}{bidirectional}

\newcommand{\ini}[1]{\ensuremath{A_{#1}}}
\newcommand{\outi}[1]{\ensuremath{B_{#1}}}

\begin{figure}[t]
\centering  
\scalebox{.8}{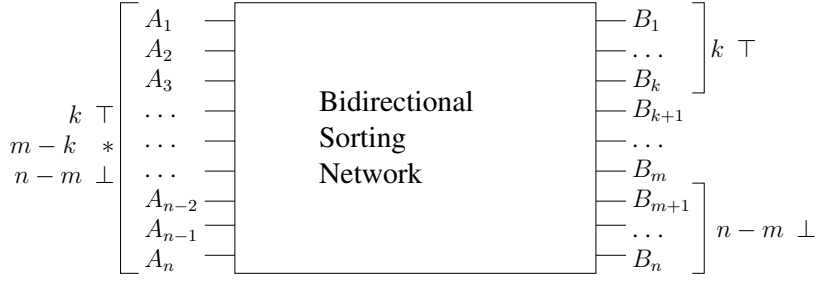}
\caption{\label{fig:snetwork} The basic schema of a bidirectional sorting network. 
}
\end{figure}

%\bcp{\mu}{\sn}{\eta}

%\PTTODO{Verificare la sezione}

%\TODO{change all indexes from $0..n-1$ to $1..n$?\\}

%\RSTODO{spiega notazione: $*$ e  \bcp{\mu}{\sn}{\eta}\\}
Notationally, the symbols $\top,\bot,*$ denote respectively ``true'',
``false'' and ``unassigned''. We represent  truth assignment as sets
(or conjunctions) of literals s.t. a positive \resp{negative} literal
denotes the fact that the corresponding atom is assigned to $\top$
\resp{$\bot$}.
Given a Boolean formula $\vi$ and two truth assignments $\mu,\eta$ on
the atoms in \vi, ``\bcp{\mu}{\vi}{\eta}'' denotes the
fact that all literals in $\eta$ are inferred by BCP on $\vi$ if
all literals in $\mu$ are asserted. (Notice that ``\bcp{\mu}{\vi}{\eta}''
is stronger than ``$\vi\wedge\mu\models\eta$''.)

When dealing with \maxsmt 
and \omt with PB objectives in the form
\begin{eqnarray}
\label{eq:obj}
\textstyle 
  \cost = w \cdot \sum_{i=1}^{n} \ini{i}
\end{eqnarray} 
\noi
a solution for improving search efficiency 
is to reduce the dependency on the expensive \laratsolver{} by better 
exploiting BCP with the aid of Boolean
{\em bidirectional sorting networks}.

\begin{definition}
\label{def:bcpcompliant}
Let \sn{} be  a CNF Boolean formula on $n$ input Boolean variables $\ins{}\defas\set{\ini1, ...,
  \ini{n}}$ and $n$ output Boolean variables $\outs{}\defas
\set{\outi1, ..., \outi{n}}$, possibly involving also
auxiliary Boolean variables which are not mentioned.

  We say that \sn{} is a  {\bf \bidirectional{}} {\bf
  sorting network} if and only if, 
for every $m$ %, $i$  
and $k$ s.t. $n\ge m %> i 
\ge k \ge 0$ and for every
partial truth assignment $\mu$ s.t. $\mu$ assigns exactly 
$k$ input variables $\ini{i}$ to
$\top$ and $n-m$ variables $\ini{i}$ to $\false$:
\begin{eqnarray}
  \label{eq:sort-bcp1}
&&  
\bcp{\mu}{\sn}{\set{\pos  \outi1,...,\pos \outi{k}}},
%&\ \ &
\\
  \label{eq:sort-bcp2}
&&  
\bcp{\mu}{\sn}{\set{\neg \outi{m+1},...,\neg \outi{n}}}.
\\
%
%% \end{eqnarray}
%% We say that a \bcpcompliant{} \sn{}  is also {\bf \bidirectional{}} iff:
% , 
% for every $m$, $k$  and $\mu$  as above and for every $i\in[k..m-1]$:
%% \ignore{%%% PROBABILMENTE NON NECESSARIE
%% \PTTODO{\eqref{eq:sort-bidirectional1}-\eqref{eq:sort-bidirectional2}
%%   sono corrette/necessarie?}}
%% \begin{eqnarray}
%% \ignoreinshort{%%% DON't WORK FOR Sinz's encoding
%% % NOTE: not mapped from 0-n-1 to 1-n
%%   \label{eq:sort-bidirectional1}
%% && \bcp{\set{\pos \outi{i}}}{\sn}{\set{\pos \outi0,...,\pos \outi{i-1}}},
%% \\
%%   \label{eq:sort-bidirectional2}
%% && \bcp{\set{\neg \outi{i}}}{\sn}{\set{\neg \outi{i+1},...,\neg \outi{n-1}}}.
%% \\
%% }
  \label{eq:sort-bidirectional3}
&& \bcp{\mu\cup\set{\neg \outi{k+1}}}{\sn}{\set{\neg \ini{i}\ s.t.\ \ini{i}\
    unassigned\ in\ \mu }},
\\
  \label{eq:sort-bidirectional4}
&& \bcp{\mu\cup\set{\pos \outi{m}}}{\sn}{\set{\ini{i}\ s.t.\ \ini{i}\
    unassigned\ in\ \mu }}.
\end{eqnarray}
\end{definition}

\ignore{%%%%%%% OLD DEFINITION
\begin{definition}
\label{def:bcpcompliant}
  We say  \sn{} is {\bf \bcpcompliant{}} if and only if, 
for every $m$ %, $i$  
and $k$ s.t. $n\ge m %> i 
\ge k \ge 0$ and for every
partial truth assignment $\mu$ s.t. $\mu$ assigns exactly 
$k$ input variables $\ini{i}$ to
$\top$ and $n-m$ variables $\ini{i}$ to $\false$:
\begin{eqnarray}
  \label{eq:sort-bcp1}
&&  
\bcp{\mu}{\sn}{\set{\pos  \outi1,...,\pos \outi{k}}},
%&\ \ &
\\
  \label{eq:sort-bcp2}
&&  
\bcp{\mu}{\sn}{\set{\neg \outi{m+1},...,\neg \outi{n}}}.
\end{eqnarray}
We say that a \bcpcompliant{} \sn{}  is also {\bf \bidirectional{}} iff:
% , 
% for every $m$, $k$  and $\mu$  as above and for every $i\in[k..m-1]$:
\ignore{%%% PROBABILMENTE NON NECESSARIE
\PTTODO{\eqref{eq:sort-bidirectional1}-\eqref{eq:sort-bidirectional2}
  sono corrette/necessarie?}}
\begin{eqnarray}
\ignoreinshort{%%% DON't WORK FOR Sinz's encoding
% NOTE: not mapped from 0-n-1 to 1-n
  \label{eq:sort-bidirectional1}
&& \bcp{\set{\pos \outi{i}}}{\sn}{\set{\pos \outi0,...,\pos \outi{i-1}}},
\\
  \label{eq:sort-bidirectional2}
&& \bcp{\set{\neg \outi{i}}}{\sn}{\set{\neg \outi{i+1},...,\neg \outi{n-1}}},
\\
}
  \label{eq:sort-bidirectional3}
&& \bcp{\mu\cup\set{\neg \outi{k+1}}}{\sn}{\set{\neg \ini{i}\ s.t.\ \ini{i}\
    unassigned\ in\ \mu }},
\\
  \label{eq:sort-bidirectional4}
&& \bcp{\mu\cup\set{\pos \outi{m}}}{\sn}{\set{\ini{i}\ s.t.\ \ini{i}\
    unassigned\ in\ \mu }}.
\end{eqnarray}
\end{definition}
} %%%%%%% OLD DEFINITION

\noi
The schema of a bidirectional sorting network is 
depicted in Figure~\ref{fig:snetwork}.

\eqref{eq:sort-bcp1}-\eqref{eq:sort-bcp2} state that 
the output values \outs{} of \sn{} are
propagated  from the inputs \ins{} via BCP. 
\ignoreinshort{%%% DON't WORK FOR Sinz's encoding
\eqref{eq:sort-bidirectional1}-\eqref{eq:sort-bidirectional2} force
\outs{} to be sorted also when output variables are assigned rather
than being propagated from the inputs.
}
\eqref{eq:sort-bidirectional3}-\eqref{eq:sort-bidirectional4} describe
how assigning output variables \outs{} propagates back to input
variables \ins{}: 
\eqref{eq:sort-bidirectional3}
states that, when $k$ $A_i$'s are true and $\outi{k+1}$ is false, 
then all other $A_i$'s are forced to be false by BCP;
dually, 
\eqref{eq:sort-bidirectional4} states that, when $n-m$ $A_i$'s are
false and $\outi{m}$ is true, 
then all other $A_i$'s are forced to be true by BCP. 
%
%\ignore{%%% PROBABILMENTE NON NECESSARIE
%}
%
% any partial assignment $\mu_0$ which does not verify {\em
%   bi-directional} either is extended by BCP to one which verifies it, or a
% conflict is produced if no such extension exists. 
% %
% For instance:
% %
% \begin{itemize}
% \item 
% if (exactly) $k$ \ini{i}s are $\top$ and $n-m$
%   are $\bot$, then $\outi0, ..., \outi{k-1}$ are set $\top$ and 
% $\outi{m}, ..., \outi{n-1}$ are set $\bot$ by BCP (if unassigned);
% \item 
% if $\outi{i}$ is 
% $\top$, then $\outi{0},...,\outi{i-1}$ are also set to $\top$
% by BCP (if unassigned);
% \item 
% if $\outi{i}$ is 
% $\bot$, then $\outi{i+1},...,\outi{n-1}$ are also set to $\top$
% by BCP (if unassigned);
% \item if $\outi{k-1}$ is $\top$ and $n-k$ \ini{i}s 
% $\ini{i}$ are  $\bot$, then by BCP all other unassigned
% $\ini{i}$s are automatically set to $\top$; 
% \item if $\outi{k+1}$ is $\bot$ and $n-k$ \ini{i}s 
% $\ini{i}$ are  $\top$, then by BCP all other unassigned
% $\ini{i}$s are automatically set to $\bot$;
% \end{itemize}
%
(If any of the above BCP assignments conflicts with some previous
assignment, a conflict is produced.)

% vice-versa, if $\outi{k+1}$ is
% forced to be False (that is, at most $k$ inputs can be True) and $n-k$
% inputs $\ini{i}$ are True, then all other unassigned $\ini{i}$s
%  are automatically set to False.

Given an \omt problem $\tuple{\varphi,\cost}$, where 
$\cost$ is as in \eqref{eq:obj}, and a Boolean formula 
$\sn$ encoding a  \bidirectional{} sorting network relation 
as in Definition~\ref{def:bcpcompliant},
we extend $\varphi$ in \eqref{eq:maxsmt2}-\eqref{eq:maxsmt3} as follows: 
\begin{equation}
\label{eq:circuitb}
\varphi' = \varphi \wedge \sn
 \wedge \bigwedge_{i=1}^{n} 
\begin{cases}
(\neg \outi{i} \vee (i\cdot w \leq \cost))\ \wedge \\
(\outi{i} \vee (\cost \leq (i - 1) \cdot w))\ \wedge \\
(\neg(i\cdot w \leq \cost) \vee \neg(\cost \leq (i - 1) \cdot w)) 
\end{cases}
\end{equation}
\noi
and optimize $\cost$ over $\varphi'$.
Notice here that the third line in equation \ref{eq:circuitb} is
\larat-valid, but it allows for implying the negation of $(\cost \leq (i - 1)
\cdot w)$ from $(i\cdot w \leq \cost)$ (and vice versa) directly by BCP,
without any call to the \laratsolver{}.

%%% END COMMENTED PART

Consider \eqref{eq:sort-bcp1}-\eqref{eq:sort-bcp2} and 
assume that $\mu$ assigns $k$ $A_i$s to $\top$ and $n-m$ to $\bot$ as in 
 Definition~\ref{def:bcpcompliant}.
%  and that $\mu$ contains already all
%  literals which can be produced via BCP on 
% \eqref{eq:maxsmt2}-\eqref{eq:maxsmt3},
% \eqref{eq:sort-bcp1}-\eqref{eq:sort-bidirectional4} and
% \eqref{eq:circuitb}.
Then
\eqref{eq:sort-bcp1} with \eqref{eq:circuitb} forces the unit-propagation of 
$\outi1,...,\outi{k}$, and then, among others, of
%$(1\cdot w\le \cost),...,$
$(k\cdot w \le \cost)$, % NOTE: there appear to have been a mistacke
												% here in the previous formulation which said
												% (0\cdot w \le \cost), as this would not derive
												% from the encoding of the formula [14]
%$\neg(\cost \le 0\cdot w),..., \neg(\cost \le (k-1)\cdot w)$;
while \eqref{eq:sort-bcp2} with \eqref{eq:circuitb} 
forces the unit-propagation of 
$\neg \outi{m+1},...,\neg \outi{n}$, and then, among others, of
$(\cost \le m\cdot w)$. This automatically restricts the range of
\cost to $[k\cdot w,m\cdot w]$, obtaining the same effect as 
\eqref{eq:maxsmt2}-\eqref{eq:maxsmt3}.

The benefits of the usage of \sn{} are due to both 
%We highlight in particular the benefits of both
\eqref{eq:sort-bidirectional3} and \eqref{eq:sort-bidirectional4}.
When the optimization
search finds a new minimum $k\cdot
w$ and a unit clause in the form $\neg(k\cdot
w \leq \cost)$ is learned (see e.g. \cite{st-ijcar12}) and
$\neg\outi{k}$ is unit-propagated on \eqref{eq:circuitb},
then  as soon as $k-1$  $\ini{i}$s are set to True,
all the remaining $n-k+1$ 
$\ini{i}$s are set to False by BCP  \eqref{eq:sort-bidirectional3} (!).
A dual case occurs when some lower-bound unit clause $\neg(\cost \leq k\cdot
w)$ is learned (e.g., in a binary-search step \cite{st-ijcar12}) and
$\outi{k+1}$ is unit-propagated on \eqref{eq:circuitb}: as soon as 
$n-k-1$ \ini{i}s are set to False, then  all the remaining $k+1$ 
\ini{i}s are set to True by BCP \eqref{eq:sort-bidirectional4}.

%\newpage
\begin{figure}[tb]
	\begin{center}
    \def\svgwidth{0.8\columnwidth}
    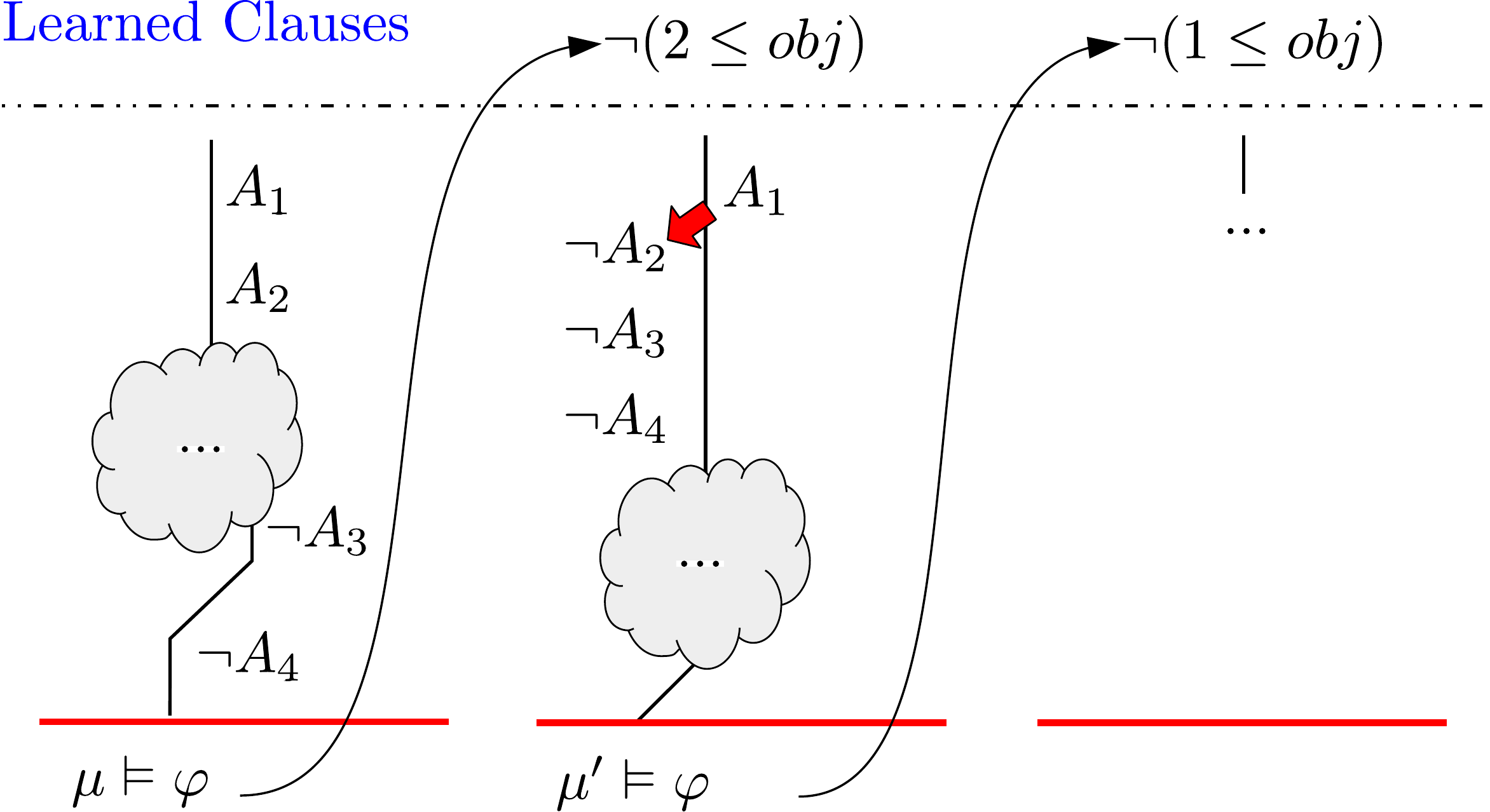
	\end{center}
\caption{\label{fig:sol1}
\textmd{An example of \omt search with sorting networks.
}}
\end{figure}
%
%{\subsubsection*{Example (continued).} 

  \begin{example}
\label{ex:withcardinalitynetwork}
  Figure~\ref{fig:sol1} considers the same scenario as in
  Example~\ref{ex:nocardinalitynetwork}, 
  in which we extend the encoding with a \bidirectional{}
  sorting-network relation as in \eqref{eq:circuitb}. 
  The behaviour is identical to that of
  Example~\ref{ex:nocardinalitynetwork}
  until the assignment $\mu$ is generated, the unit clause $\neg(2
  \leq \cost)$, and the procedure backtracks for the first
  time (Figure~\ref{fig:sol1} left). 
This causes the unit-propagation of $\neg \outi2$
  on \eqref{eq:circuitb}.
As soon as $\ini1$ is picked as new
  decision, $\neg \ini2, \neg \ini3, \neg\ini4$ are unit propagated 
\eqref{eq:sort-bidirectional3}, saving up to ${{4}\choose{2}}$
(expensive) calls
  to  the \laratsolver{} 
(Figure~\ref{fig:sol1} center).    
Then $\neg(1
  \leq \cost)$ is learned, and the search proceeds
  (Figure~\ref{fig:sol1} right).  
\hfill $\diamond$  
\end{example}

We generalize this approach %based on sorting networks 
to deal with the general 
objectives as in \eqref{eq:gen1}-\eqref{eq:gen2}.  In this case
a separate sorting circuit is generated for each term $\tau_j$, and
%some %(optional) 
constraints in the form 
\begin{eqnarray}
\textstyle
\bigwedge_{j=1}^m
\bigwedge_{i=1}^{k_j}
(\ \neg(w_j \cdot i \leq \cost)
\imp \neg(w_j \cdot i \leq \tau_j)\ ),%\  i\ \in [1, k_j],  
\end{eqnarray}
are added to ensure that the circuit is activated by BCP.
% , in the
% same spirit as the constraints in \eqref{eq:extraconstraints}. 

%\newpage
\subsection{Bidirectional Sorting Networks}
\label{sec:combining-sortingnetworks}

Unlike the usage of sorting networks in other contexts, 
which consider only \eqref{eq:sort-bcp1} and 
\eqref{eq:sort-bidirectional3} as relevant properties (e.g. \cite{Sinz05}), we are
interested in sorting networks which propagate both $\top$ and $\bot$ 
values in both
directions (i.e., which comply with all properties
  \eqref{eq:sort-bcp1}-\eqref{eq:sort-bidirectional4}). 
%
%% in all properties
%% \eqref{eq:sort-bcp1}-\eqref{eq:sort-bidirectional4}. 
To this extent, 
we have considered two encodings: 
the sequential counter encoding in \cite{Sinz05}, which we have 
extended to comply with all properties  \eqref{eq:sort-bcp1}-\eqref{eq:sort-bidirectional4}, 
and the cardinality network encoding in \cite{Asin2011,parametricCardinalityConstraint}. 
\ignore{
Notice that, in contrast with some of the literature which focuses on
only one direction of cardinality constraints 
\eqref{eq:sort-bcp1}-\eqref{eq:sort-bcp2} (e.g. \cite{Sinz05}), in our context we are
interested in a bidirectional encodings, so that to
ensure the backward propagation of the $\outi{k}$ values 
\eqref{eq:sort-bidirectional3}-\eqref{eq:sort-bidirectional4}.
}
%\RSTODO{RIVEDI QUESTA SOTTOSEZIONE}

\ignore{%%% SECONDO ME GIA? SPIAGATO
This is due to
the fact that our \omt solver applies this technique to deal not only
with MaxSMT problems but also with Pseudo-Boolean objectives that can
be either minimized or maximized, and other Pseudo-Boolean sums
subject to simple cardinality constraints outside of an optimization
context.
}

%\newpage
\paragraph{Bidirectional Sequential Counter Encoding.}

The sequential counter encoding $LT_{SEQ}^{n,k}$ for $\leq k (\ini1,
..., \ini{n})$ presented in \cite{Sinz05} consists of 
$O(k\cdot n)$ clauses and variables and complies with
\eqref{eq:sort-bcp1} and \eqref{eq:sort-bidirectional3}. % and is arc-consistent. 
The circuit is given by the composition of $n$ sub-circuits, each of which
computes $S_i = \sum_{j=1}^{i}\ini{j}$, represented in unary form with
the bits $S_{i, j}${, i.e., $S_{i,j}=\top$ if $ \sum_{r=1}^{i}A_{r}\ge j$,
so that $\outi{j}\defas S_{n,j}$, $j\in[1..n]$.} 
{The (CNF version of
  the)\footnote{Here \eqref{eq:sinz-encoding1}-\eqref{eq:sinz-reverse-encoding2}
  are written as implications to emphasize the directionality of the
  encodings.}} following formula is the %original %propositional  
encoding of $LT_{SEQ}^{n,k}$ presented in \cite{Sinz05}, with $k\defas
n$:
%\RSTODO{merge le due versioni in una sola ``$\iff$''?}
\ignore{%%%% VERSIONE CNF
\begin{eqnarray}
\label{eq:sinz-encoding1}
\textstyle
(\neg \ini{1}\vee S_{1,1}) \wedge
\bigwedge_{i=2}^{n}
	\{
	(\neg \ini{i}\vee S_{i,1}) \wedge
	(\neg S_{i-1,1}\vee S_{i,1}) \wedge
	(\neg \ini{i}\vee\neg S_{i-1,n}) 
	\} \wedge \\
\nonumber
\textstyle
\bigwedge_{j=2}^{n} \{
	(\neg S_{1,j}) \} \wedge
\bigwedge_{i,j=2}^{n} \{
	(\neg \ini{i}\vee\neg S_{i-1,j-1}\vee S_{i,j}) \wedge
	(\neg S_{i-1,j}\vee S_{i,j}) \}
\end{eqnarray}
}
%%%% VERSIONE "->"
\begin{eqnarray}
\label{eq:sinz-encoding1}
\textstyle
(\ini{1}\imp S_{1,1}) \wedge
\bigwedge_{i=2}^{n}
	\{
	((\ini{i}\vee S_{i-1,1})\imp S_{i,1}) 
        \} \wedge \\
\label{eq:sinz-encoding2}
\textstyle
\bigwedge_{i=2}^{n}
	\{
%	(S_{i-1,1}\imp S_{i,1}) \wedge
	(\neg\ini{i}\vee\neg S_{i-1,n}) 
	\} \wedge 
%\nonumber
\textstyle
\bigwedge_{j=2}^{n} \{
	(\neg S_{1,j}) \} \wedge
\\
\label{eq:sinz-encoding3}
%\nonumber
\textstyle
\bigwedge_{i,j=2}^{n} \{
	(((\ini{i}\wedge S_{i-1,j-1})\vee S_{i-1,j})\imp S_{i,j}) 
%\wedge	(S_{i-1,j}\imp S_{i,j}) 
\}
\end{eqnarray}
Notice that, in order to reduce the size of the encoding,
 in {\eqref{eq:sinz-encoding1}-\eqref{eq:sinz-encoding3}
 only right implications ``$\imp$'' were used to encode each
gate in the Boolean sorting circuit 
\cite{Sinz05}, so that 
%were obtained from a sorting circuit by encoding each gate 
%
%For our own application, we need both directions of the equations to ensure 
%that  the $\outi{k}$ values are correctly propagated even when the 
%circuit is subject to maximization constraints or no constraint at all.
%
%Thus, we strengthened the formula given in \cite{Sinz05} with the
%following clauses:
% {%
% (Here 
% $S_{i-1,j}=\top$ if $ \sum_{r=0}^{i}A_{r}>j$, 
% so that $\outi{j}\defas S_{n-1,j}$, $j\in[0..n-1]$.)
% }
%
\eqref{eq:sinz-encoding1}-\eqref{eq:sinz-encoding3} 
complies with
\eqref{eq:sort-bcp1} and \eqref{eq:sort-bidirectional3} 
but not with \eqref{eq:sort-bcp2} and
\eqref{eq:sort-bidirectional4}.
 %\RSTODO{verifica}.
To cope with this fact,} we have added the
following part, which 
reintroduces the left implications ``$\limp$'' of the encoding
of each gate in
\eqref{eq:sinz-encoding1} and \eqref{eq:sinz-encoding3}, 
making  it compliant also with \eqref{eq:sort-bcp2} and
\eqref{eq:sort-bidirectional4}: 
\ignore{%%%% CNF VERSION
\begin{eqnarray}
\textstyle
(\ini{1}\vee\neg S_{1,1}) \wedge
\bigwedge_{i=2}^{n}
\{
	(\neg S_{i,1}\vee \ini{i}\vee S_{i-1,1})
\} \wedge\\
\nonumber
\textstyle
\bigwedge_{i,j=2}^{n}
(\neg S_{i,j}\vee S_{i-1,j}\vee(S_{i-1,j-1}\wedge \ini{i})).
\end{eqnarray}
}
%%% "-> version
\begin{eqnarray}
\label{eq:sinz-reverse-encoding1}
\textstyle
(\ini{1}\limp S_{1,1}) \wedge
\bigwedge_{i=2}^{n}
\{
        ((\ini{i}\vee S_{i-1,1})\limp S_{i,1})
%	(\neg S_{i,1}\vee \ini{i}\vee S_{i-1,1})
\} \wedge\\
%\nonumber
\label{eq:sinz-reverse-encoding2}
\textstyle
\bigwedge_{i,j=2}^{n}
(((\ini{i}\wedge S_{i-1,j-1})\vee S_{i-1,j}) \limp S_{i,j}).
\end{eqnarray}

\ignore{ % deprecated formula
% NOTE: not translated from 0-n-1 to 1-n
\[
\begin{array}{l}
(\ini{0}\vee\neg S_{0,0})\\
\begin{rcases*}
(\neg S_{i,0}\vee \ini{i}\vee S_{i-1,0})\:\:\:\:\:\:\:\:\:\:\:\:\:\:\:\:\:\:\:\:\:\:\:\:\:\:\:\:\:\\
\end{rcases*}
i \in [1,n[ \\
\begin{rcases*}
(\neg S_{i,j}\vee S_{i-1,j}\vee(S_{i-1,j-1}\wedge \ini{i}))\:\:\:\:\:\\
\end{rcases*} 
j \in [1,n[ \: \wedge \: i \in [1,n[ \\
\end{array}
\].
}

\paragraph{Bidirectional Cardinality Network Encoding.}
The cardinality network encoding presented in
\cite{EenS06,Asin2011,parametricCardinalityConstraint}, based on the
underlying sorting scheme of the well-known \textit{merge-sort}
algorithm, has complexity $O(n \log^2 k)$ in the number of clauses and
variables.
Due to space limitations, we refer the reader to
\cite{parametricCardinalityConstraint,Asin2011} for the encoding of cardinality
networks we used in our own work. Notice that, differently than in the
previous case, this sorting network propagates both $\top$ and $\bot$ 
values in both
directions (i.e., it complies with all properties
  \eqref{eq:sort-bcp1}-\eqref{eq:sort-bidirectional4} \cite{Asin2011,parametricCardinalityConstraint} and it is thus
suitable to be used within \omt without modifications.

%% file: circuit.pdf_t
\begin{picture}(0,0)%
\includegraphics{circuit.pdf}%
\end{picture}%
\setlength{\unitlength}{4144sp}%
\begingroup\makeatletter\ifx\SetFigFont\undefined%
\gdef\SetFigFont#1#2#3#4#5{%
  \reset@font\fontsize{#1}{#2pt}%
  \fontfamily{#3}\fontseries{#4}\fontshape{#5}%
  \selectfont}%
\fi\endgroup%
\begin{picture}(4530,2056)(1336,-2555)
\put(1576,-691){\makebox(0,0)[lb]{\smash{{\SetFigFont{12}{14.4}{\familydefault}{\mddefault}{\updefault}{\color[rgb]{0,0,0}\ini{1}}%
}}}}
\put(1576,-916){\makebox(0,0)[lb]{\smash{{\SetFigFont{12}{14.4}{\familydefault}{\mddefault}{\updefault}{\color[rgb]{0,0,0}\ini{2}}%
}}}}
\put(1576,-1141){\makebox(0,0)[lb]{\smash{{\SetFigFont{12}{14.4}{\familydefault}{\mddefault}{\updefault}{\color[rgb]{0,0,0}\ini{3}}%
}}}}
\put(1576,-2041){\makebox(0,0)[lb]{\smash{{\SetFigFont{12}{14.4}{\familydefault}{\mddefault}{\updefault}{\color[rgb]{0,0,0}\ini{n-2}}%
}}}}
\put(1576,-2266){\makebox(0,0)[lb]{\smash{{\SetFigFont{12}{14.4}{\familydefault}{\mddefault}{\updefault}{\color[rgb]{0,0,0}\ini{n-1}}%
}}}}
\put(1576,-2491){\makebox(0,0)[lb]{\smash{{\SetFigFont{12}{14.4}{\familydefault}{\mddefault}{\updefault}{\color[rgb]{0,0,0}\ini{n}}%
}}}}
\put(1576,-1366){\makebox(0,0)[lb]{\smash{{\SetFigFont{12}{14.4}{\familydefault}{\mddefault}{\updefault}{\color[rgb]{0,0,0}. . .}%
}}}}
\put(1576,-1591){\makebox(0,0)[lb]{\smash{{\SetFigFont{12}{14.4}{\familydefault}{\mddefault}{\updefault}{\color[rgb]{0,0,0}. . .}%
}}}}
\put(1576,-1816){\makebox(0,0)[lb]{\smash{{\SetFigFont{12}{14.4}{\familydefault}{\mddefault}{\updefault}{\color[rgb]{0,0,0}. . .}%
}}}}
\put(2881,-1591){\makebox(0,0)[lb]{\smash{{\SetFigFont{14}{16.8}{\familydefault}{\mddefault}{\updefault}{\color[rgb]{0,0,0}Sorting}%
}}}}
\put(2881,-1861){\makebox(0,0)[lb]{\smash{{\SetFigFont{14}{16.8}{\familydefault}{\mddefault}{\updefault}{\color[rgb]{0,0,0}Network}%
}}}}
\put(2881,-1321){\makebox(0,0)[lb]{\smash{{\SetFigFont{14}{16.8}{\familydefault}{\mddefault}{\updefault}{\color[rgb]{0,0,0}Bidirectional}%
}}}}
\put(5221,-1591){\makebox(0,0)[lb]{\smash{{\SetFigFont{12}{14.4}{\familydefault}{\mddefault}{\updefault}{\color[rgb]{0,0,0}. . .}%
}}}}
\put(5221,-916){\makebox(0,0)[lb]{\smash{{\SetFigFont{12}{14.4}{\familydefault}{\mddefault}{\updefault}{\color[rgb]{0,0,0}. . .}%
}}}}
\put(5221,-2266){\makebox(0,0)[lb]{\smash{{\SetFigFont{12}{14.4}{\familydefault}{\mddefault}{\updefault}{\color[rgb]{0,0,0}. . .}%
}}}}
\put(1351,-1411){\makebox(0,0)[rb]{\smash{{\SetFigFont{12}{14.4}{\familydefault}{\mddefault}{\updefault}{\color[rgb]{0,0,0}$k$\ \ $\top$}%
}}}}
\put(1351,-1636){\makebox(0,0)[rb]{\smash{{\SetFigFont{12}{14.4}{\familydefault}{\mddefault}{\updefault}{\color[rgb]{0,0,0}$m-k$\ \ \ \ $*$}%
}}}}
\put(1351,-1861){\makebox(0,0)[rb]{\smash{{\SetFigFont{12}{14.4}{\familydefault}{\mddefault}{\updefault}{\color[rgb]{0,0,0}$n-m$\ \ $\bot$}%
}}}}
\put(5221,-1141){\makebox(0,0)[lb]{\smash{{\SetFigFont{12}{14.4}{\familydefault}{\mddefault}{\updefault}{\color[rgb]{0,0,0}\outi{k}}%
}}}}
\put(5221,-1366){\makebox(0,0)[lb]{\smash{{\SetFigFont{12}{14.4}{\familydefault}{\mddefault}{\updefault}{\color[rgb]{0,0,0}\outi{k+1}}%
}}}}
\put(5221,-1816){\makebox(0,0)[lb]{\smash{{\SetFigFont{12}{14.4}{\familydefault}{\mddefault}{\updefault}{\color[rgb]{0,0,0}\outi{m}}%
}}}}
\put(5221,-2041){\makebox(0,0)[lb]{\smash{{\SetFigFont{12}{14.4}{\familydefault}{\mddefault}{\updefault}{\color[rgb]{0,0,0}\outi{m+1}}%
}}}}
\put(5221,-2491){\makebox(0,0)[lb]{\smash{{\SetFigFont{12}{14.4}{\familydefault}{\mddefault}{\updefault}{\color[rgb]{0,0,0}\outi{n}}%
}}}}
\put(5221,-691){\makebox(0,0)[lb]{\smash{{\SetFigFont{12}{14.4}{\familydefault}{\mddefault}{\updefault}{\color[rgb]{0,0,0}\outi{1}}%
}}}}
\put(5806,-916){\makebox(0,0)[lb]{\smash{{\SetFigFont{12}{14.4}{\familydefault}{\mddefault}{\updefault}{\color[rgb]{0,0,0}$k$\ \ $\top$}%
}}}}
\put(5851,-2266){\makebox(0,0)[lb]{\smash{{\SetFigFont{12}{14.4}{\familydefault}{\mddefault}{\updefault}{\color[rgb]{0,0,0}$n-m$\ \ $\bot$}%
}}}}
\end{picture}%

%% file: png/2.pdf_tex
%% Creator: Inkscape inkscape 0.91, www.inkscape.org
%% PDF/EPS/PS + LaTeX output extension by Johan Engelen, 2010
%% Accompanies image file '2.pdf' (pdf, eps, ps)
%%
%% To include the image in your LaTeX document, write
%%   \input{<filename>.pdf_tex}
%%  instead of
%%   \includegraphics{<filename>.pdf}
%% To scale the image, write
%%   \def\svgwidth{<desired width>}
%%   \input{<filename>.pdf_tex}
%%  instead of
%%   \includegraphics[width=<desired width>]{<filename>.pdf}
%%
%% Images with a different path to the parent latex file can
%% be accessed with the `import' package (which may need to be
%% installed) using
%%   \usepackage{import}
%% in the preamble, and then including the image with
%%   \import{<path to file>}{<filename>.pdf_tex}
%% Alternatively, one can specify
%%   \graphicspath{{<path to file>/}}
%% 
%% For more information, please see info/svg-inkscape on CTAN:
%%   http://tug.ctan.org/tex-archive/info/svg-inkscape
%%
\begingroup%
  \makeatletter%
  \providecommand\color[2][]{%
    \errmessage{(Inkscape) Color is used for the text in Inkscape, but the package 'color.sty' is not loaded}%
    \renewcommand\color[2][]{}%
  }%
  \providecommand\transparent[1]{%
    \errmessage{(Inkscape) Transparency is used (non-zero) for the text in Inkscape, but the package 'transparent.sty' is not loaded}%
    \renewcommand\transparent[1]{}%
  }%
  \providecommand\rotatebox[2]{#2}%
  \ifx\svgwidth\undefined%
    \setlength{\unitlength}{674.78743453bp}%
    \ifx\svgscale\undefined%
      \relax%
    \else%
      \setlength{\unitlength}{\unitlength * \real{\svgscale}}%
    \fi%
  \else%
    \setlength{\unitlength}{\svgwidth}%
  \fi%
  \global\let\svgwidth\undefined%
  \global\let\svgscale\undefined%
  \makeatother%
  \begin{picture}(1,0.54509557)%
    \put(0,0){\includegraphics[width=\unitlength,page=1]{png/2.pdf}}%
  \end{picture}%
\endgroup%

%% file: combining_trail.tex
\vspace{0.5em} \noindent{} Both of the previous encodings are
istantiated assuming $k = n$, since the sorting network is generated
prior to starting the search. Therefore, the cardinality network
circuit looks more appealing than the sequential counter encoding due
to its lower complexity in terms of clauses and variables employed.

%% file: expeval.tex
%\TODO{Eliminare tutto il pippone su assert-soft?}
We extended \optimathsat with a novel internal 
preprocessing step, 
\ignoreinshort{ for dealing with \texttt{assert-soft} constraints
 (see \cite{st_cav15,optimathsat-url})}
which automatically augments the
input formula with a sorting network circuit of choice between the
bidirectional sequential counter and the cardinality network, as
described in \sref{sec:combining}. 
\ignore{
We recall here that, in \optimathsat, the \texttt{assert-soft}
statement can be used not only to encode \maxsmt problems like in
\nuZ, but also generic Pseudo-Boolean terms that can be combined with
other constraints or objective functions \cite{st_tacas15,st_cav15}.
}
To complete our comparison, we also implemented in \optimathsat{}
two \maxsat-based approaches, 
the max-resolution approach implemented
in \nuz \cite{narodytskab14,bjorner_scss14} and
(for \maxsmt only)
the lemma-lifting approach of
\cite{cgss_sat13_maxsmt}, using \maxino{} \cite{alviano_ijcai15}  as external
\maxsat solver. 

Here we present an extensive empirical evaluation of various
\maxsat-based and \omt-based techniques in \optimathsat{}
\cite{st_cav15,optimathsat-url} and \nuz
\cite{bjorner_tacas15,nuz-url}.
{Overall, we considered $>\!20,000$  OMT problems and run
$>\!270,000$ job pairs.
The problems were
produced either by {\sc CGM-Tool}~\cite{cgm-tool} from optimization of
Constrained Goal Models~\cite{nguyensgm_er16,nguyensgm16} (a modeling
and automated-reasoning tool for requirement engineering) 
or by  {\sc PyLMT} \cite{pylmt_url} from 
(Machine) 
Learning Modulo Theories \cite{teso_aij15}.
We partition these problems into two distinct categories.
%
% \begin{itemize}
% \item 
%
In \sref{sec:expeval_suitable} we analyze problems which are {\em solvable 
%directly or indirectly  
by \maxsat-based approaches},
like those with PB objective functions or their lexicographic combination,
so that to allow both \nuz and \optimathsat{} to use their \maxsat-specific 
max-resolution engines (plus others).   
In \sref{sec:expeval_unsuitable} we analyze problems which {\em cannot} be
solved by \maxsat-based approaches, because the objective functions
involve some non-PB components, forcing to restrict 
 to OMT-based approaches only.
%
%\end{itemize}
%
% \RSTODO{ Goal: twofold: 
% 1) compare \maxsat-based and OMT-based approaches when the former is applicable
% 2) verify the benefits of sorting 
% network on OMT-based approaches.}

%\begin{rschange}
The goal of this empirical evaluation is manyfold: 
\begin{renumerate}
\item
 compare the performance of \maxsat-based approaches
wrt. \omt-based ones, on the kind of OMT problems where the former are
applicable; 
\item 
evaluate the benefits of sorting-network encodings 
with \omt-based approaches (and also with \maxsat-based ones);
\item
compare the performances of \optimathsat{} with those of \nuz{}.
% on the various configurations. 
\end{renumerate}

%  (in particular, 
% evaluate whether the speed-up obtained, if any, outweighs the
% cost of its generation). 
%\end{renumerate}

% \RSTODO{2 kinds of problems: solvable and unsolvable with
% \maxsat-based approach.}

% In the following experiments we
% evaluate whether the speed-up obtained by extending the input formula
% with a sorting network during the optimization search can outweigh the
% cost of its generation step within the \omt solver itself, as opposed
% to demanding the end-user to do it offline.

\ignore{
%
% NOTE: here I address both of the PTTODO{} suggestions below in a single sweep.
%
  Due to time limitations and to the large amount of problems used
  ($>20,000$), in all our experiments we performed a
  partial correctness verification, by checking that the different
  solvers, and their various configurations, agreed on the optimal
  value of every objective in a formula. Whenever this was not the
  case, a deeper correctness check was performed using the \zthree
  \smt solver. First, we check the satisfiability of the optimum model
  conjoined with the original formula. Then, we use the \smt solver to
  verify the absence of improving solutions over the optimal objective
  function value.  }

%We considered the following solvers configurations for teh .
For goals (i)  and (ii) we used the following configurations of 
\optimathsat. 
\ignoreinshort{
(If not specified otherwise, penalties/PB objectives
are encoded via  {\tt assert-soft}.)~
}
%\footnote{}
\begin{itemize}
\item[\omt-based:] standard, enriched with the bidirectional
sequential-counter  and cardinality sorting network;
\item[\maxsat-based:] the above-mentioned max-resolution
implementation, with and without the cardinality sorting network, 
and lemma-lifting (for pure \maxsmt only).
\end{itemize}
\noindent
For goal (iii) we also used the following configurations of \nuz.~%
\footnote{%
Notice that, unlike \optimathsat,  \nuz selects automatically its presumably-best
configuration for a given input problem.
In particular, when \maxsmt-encodable problems are fed to \nuz{} --like, e.g., those
in \sref{sec:expeval_suitable}-- 
\nuz forces automatically the choice of the \maxsat-based configuration, 
allowing the user only the choice of the \maxsat algorithm.
Thus we could not test \nuz{} also with \omt-based configuration for 
the problems in \sref{sec:expeval_suitable}. Alternatively, we should
have disguised the input problem, with the risk of affecting the
significance or the result.
}
\ignoreinshort{(\nuz allows {\tt assert-soft} only with \maxsmt-encodable
problems (e.g., \sref{sec:expeval_suitable}),
forcing automatically a \maxsat-based engine in such case, 
allowing only the choice of the \maxsat algorithm.)\\
(\nuz recognizes automatically \maxsmt-encodable
problems --like, e.g., those in \sref{sec:expeval_suitable}: if so, 
it forces automatically the choice of a \maxsat-based engine, 
allowing the user only the choice of the \maxsat algorithm.)
}
\begin{itemize}
\item[\omt-based:] standard (encoded as in \eqref{eq:maxsmt2}-\eqref{eq:maxsmt3}).
\item[\maxsat-based:] %(encoded via {\tt assert-soft}) 
using alternatively the internal implementations of the 
max-resolution \cite{narodytskab14,bjorner_scss14} and {\sc Wmax}
\cite{nieuwenhuis_sat06} procedures. 
\end{itemize}

%\end{rschange}

\ignore{
In this experiment we considered the following solvers and their configurations:
\begin{itemize}
\item \vZ with the \maxsat engine based on max-resolution
\cite{narodytskab14,bjorner_scss14} and WMax
\item \optimathsat executed either with the its novel implementation of a Maximum Resolution
	\maxsat engine or with its standard optimization search with and without cardinality 
	network constraints
\item A fresh re-implementation of the lemma lifting approach proposed in \cite{cgss_sat13_maxsmt}
	into \optimathsat{}, coupled with the \maxino \maxsat solver
\end{itemize}
}

Each job pair was run on one of two
identical Ubuntu Linux machines featuring \textit{8-core
Intel-Xeon@2.20GHz} CPU, $64$ GB of ram and kernel 3.8-0-29.
Importantly, we verified that
all tools/configurations under test
agreed on the results on all problems when terminating within the
timeout.
(The timeout varies with the benchmark sets, see
\sref{sec:expeval_suitable} and \sref{sec:expeval_unsuitable}.)
All benchmarks, as well as our experimental results and all the tools
which are
necessary to reproduce the results, are 
available \cite{tacas17-experiments-url}.
% at
%   \url{http://disi.unitn.it/trentin/resources/tacas17.tar.gz} }

%% file: expeval_lexicographic.tex
%%%%%%%%%%%%%%%%%%%%%%%%%%%%%%%%%%%%%%%%%%%%%%%%%%%%%%%%%%%%%%%%%%%%%%%%%%%%%%%%%%%%%%%%%%%%%%%
%%%%%%%%%%%%%%%%%%%%%%%%%%%%%%%%%%%%%%%%%%%%%%%%%%%%%%%%%%%%%%%%%%%%%%%%%%%%%%%%%%%%%%%%%%%%%%%
%%%%%%%%%%%%%%%%%%%%%%%%%%%%%%%%%%%%%%%%%%%%%%%%%%%%%%%%%%%%%%%%%%%%%%%%%%%%%%%%%%%%%%%%%%%%%%%
%

\input{figure-lexicographic}

%\begin{rschange}
In our first experiment we consider the set of all problems produced by 
{\sc CGM-Tool}~\cite{cgm-tool} in the
experimental evaluation in \cite{nguyensgm16}.
They consist of $18996$ automatically-generated formulas which encode
the problem of computing the lexicographically-optimum realization of a constrained goal
model \cite{nguyensgm16}, according to a prioritized list of 
(up to) three objectives \tuple{\cost_1,\cost_2,\cost_3}. 
%
% Each \omt problem contains a combination of up to three \maxsmt
% objectives sorted with lexicographic priority. 
%\footnote{
A solution {\em optimizes lexicographically}
\tuple{obj_1,...,obj_k} if it optimizes $obj_1$ and, if more than 
one such $obj_1$-optimum solutions exists, it also optimizes
$obj_2$,..., and so on;
both \omt-based and \maxsat-based techniques 
handle lexicographic optimization, by optimizing $\cost_1,\cost_2,...$ in order, fixing the value of each $obj_i$ to its optimum as soon as
it is found
 \cite{bjorner_scss14,bjorner_tacas15,st_tacas15,st_cav15}.
%}
In this experiment, we
set the timeout at $100$ seconds.
%\PTTODO{Spiegare meglio quest'ultima frase}
The results are reported in Figure~\ref{tab:maxsmt1} (top and
middle). 
%\end{rschange}

As far as \optimathsat{} (\omt-based) is concerned,
extending the input formula with either of the sorting 
networks increases the number of benchmarks solved within the timeout. 
Notably, the cardinality network encoding --which has the lowest
complexity-- scores the best both in terms of  
number of solved benchmarks and solving time.
On the other hand, the sequential counter network is affected by a
significant performance hit on a number of benchmarks, as it is
witnessed by the left scatter plot in figure
\ref{tab:maxsmt1}. This not only affects unsatisfiable benchmarks, for
which using sorting networks appears to be not beneficial in general,
but also satisfiable ones.

A possible strategy for overcoming this performance issue is to reduce
the memory footprint determined by the generation of the sorting
network circuit. This can be easily achieved by splitting each
Pseudo-Boolean sum in smaller sized chunks and generating a separate
sorting circuit for each splice.
The result of applying this enhancement, using chunks of increasing
size, is shown in Figure~\ref{tab:maxsmt1} (bottom). The data
suggest that the sequential counter encoding can 
benefit from this simple heuristic, but it does not reach the
performances of the cardinality network, which are not affected 
by this strategy. (In next experiments this strategy will be no more considered.)

{As far as \optimathsat{} (\maxsat-based) is concerned, 
we notice that it significantly outperforms all \omt-based techniques.
Remarkably, extending the input formula with the sorting networks
improves the performance also of this configuration.} 

{As far as \nuz{} (\maxsat-based) is concerned, 
we notice that when using the max-resolution \maxsat algorithm it 
outperforms all other techniques by solving all problems, 
whereas when using the \texttt{Wmax}
engine the performances decrease
drastically.
} 

%% file: figure-lexicographic.tex
\begin{figure}[t]
\centering
% \begin{subfigure}[t]
% \centering
\begin{tabular}{|l|rrr|r|}
\hline
\textbf{tool, configuration \& encoding}    & \textbf{inst.} & \textbf{term.} & \textbf{timeout} & \textbf{time (s.)} \\
\hline
%\multicolumn{5}{|l|}{\optimathsat}                         \\
\optimathsat{} (\omt-based)   & 18996    & 16316              & 2680       & 48832     \\
\optimathsat{} (\omt-based + seq. counter)   & 18996    & 16929              & 2067    & 90080     \\
\optimathsat{} (\omt-based + card. network) & 18996    & \textbf{17191}     & 1805    & 39215    \\
\hline
%\multicolumn{5}{|l|}{\optimathsat + Maximum Resolution}  \\
\optimathsat{} (\maxsat-based w. maxres)   & 18996 & 17933 & 1063 & 24369 \\
\optimathsat{} (\maxsat-based w. maxres + seq. counter) & 18996 & 18180 & 816 & 49088 \\
\optimathsat{} (\maxsat-based w. maxres + card. netw.)  & 18996 & \textbf{18197} &  799 & 26489 \\
\hline
\hline
% NEW TACAS17:
\nuZ (\maxsat-based w. maxres)       & 18996    & \blue{\textbf{18996}}              &    0    & 1640 \\
\nuZ(\maxsat-based w. wmax)         & 18996    & 16640              & 2356    & 38945 \\
%
% FROM SMT16:
%
%\nuZ(maxres)       & 18996    & 18996     & 255        & 1766     \\
%\nuZ(wmax)         & 18996    & 16650     & 3785       & 38040     \\
%
%\textbf{encoding}    & \textbf{\#total} & \textbf{\#solved} & \textbf{\#timeout} & \textbf{time (s.)} \\
%simple \omt enc. & 18996    & 16316     & 2680       & 48832     \\
%seq. counter enc. & 18996    & 16929     & 2067       & 90080     \\
%card. network enc. & 18996    & \textbf{17191}     & 1805       & 39215    \\
\hline
\end{tabular}\\
\begin{tabular}{cc}%
\hspace{-2em}
		\includegraphics[scale=0.36]{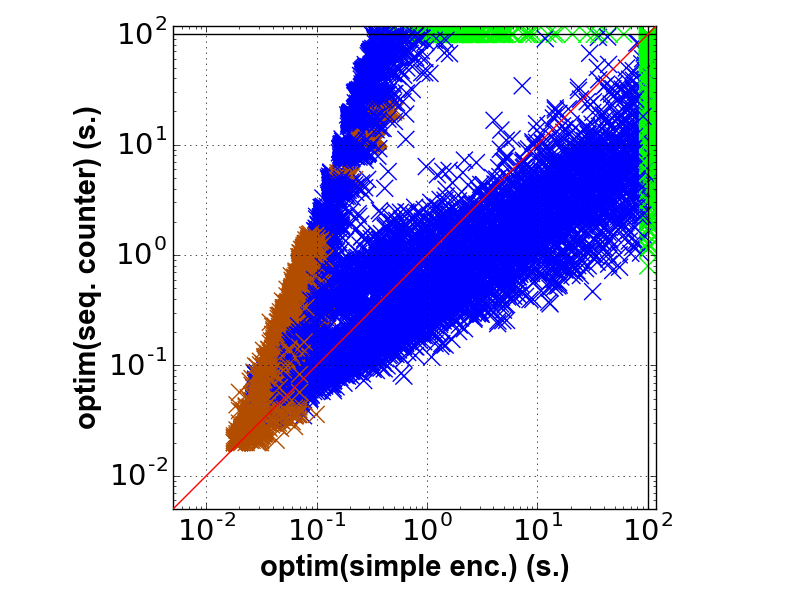}
&
\hspace{-3em}
		\includegraphics[scale=0.36]{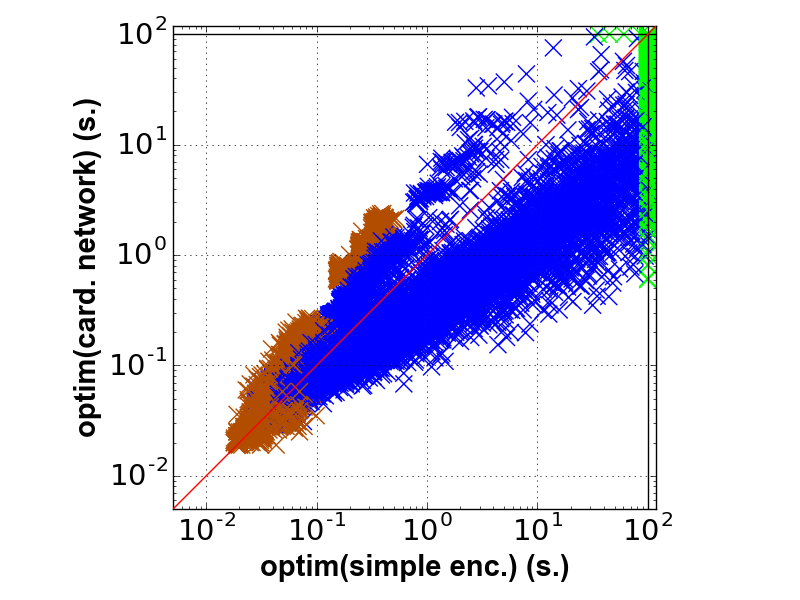}
\end{tabular}
%\end{subfigure}
%
%\begin{subfigure}[t]
\centering
\begin{tabular}{ccc}%
%\hspace{-3.5em}
\begin{tabular}{|l|rrr|r|}
\hline
\multicolumn{5}{|l|}{\textbf{partitioned sequential-counter encoding}} \\
\hline
\textbf{vars}    & \textbf{inst.} & \textbf{term.} & \textbf{timeout} & \textbf{time (s.)} \\
\hline
$\infty$  & 18996    & 16929     & 2067    & 90080     \\
10    & 18996    & 17033     & 1963    & 39035     \\
15   & 18996    & 17061     & 1935    & 39264     \\
20    & 18996    & \textbf{17152}     & 1844    & 43730     \\
\hline
\end{tabular}
& \ \ \ \ & 
%\hspace{-6.5em}
\begin{tabular}{|l|rrr|r|}
\hline
\multicolumn{5}{|l|}{\textbf{partitioned cardinality-network encoding}} \\
\hline
\textbf{vars}   & \textbf{inst.} & \textbf{term.} & \textbf{timeout} & \textbf{time (s.)} \\
\hline
$\infty$  & 18996    & \textbf{17191}     & 1805    & 39215    \\
10 & 18996    & 17058     & 1938    & 36636     \\
15 & 18996    & 17133     & 1863    & 37246     \\
20 & 18996    & 17190     & 1806    & 39492     \\
\hline
\end{tabular}
\\
\ignore{
\hspace{-2.5em}
		\includegraphics[scale=0.36]{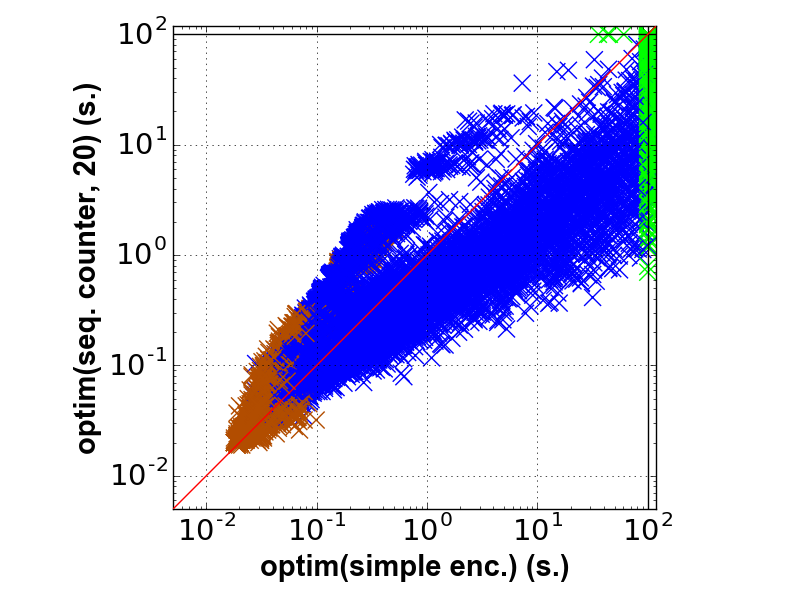}
&
\hspace{-4.5em}
		\includegraphics[scale=0.36]{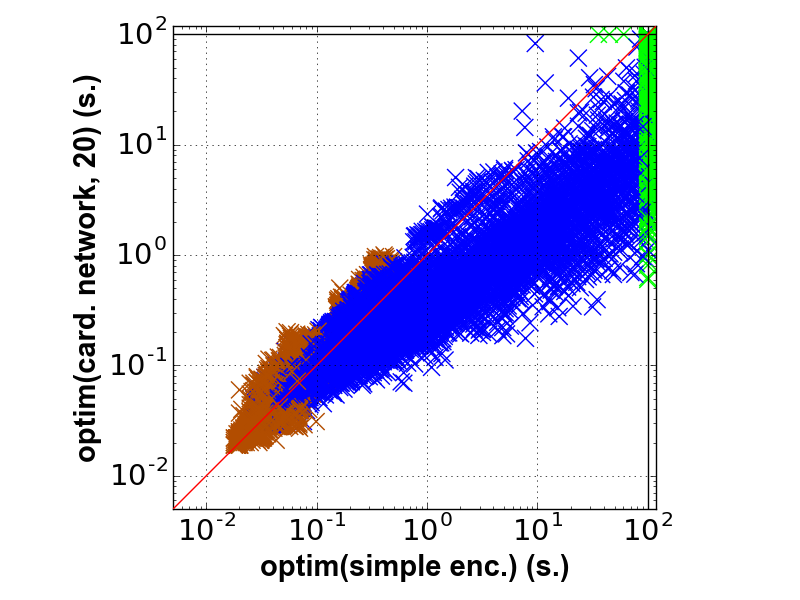}\\
}
\end{tabular} 
\caption{
\label{tab:maxsmt1}
\label{tab:maxsmt2}
[Top, table] Results of various solvers, configurations and encodings
on all the problems encoding CGM optimization with lexicographic PB
optimization of \cite{nguyensgm_er16,nguyensgm16}.
(Values in \textbf{boldface} denote the best performance of each
category; values in \textbf{\blue{blue}} denote the absolute best performance.)
\newline
[Middle, scatterplots]. Pairwise comparison on \optimathsat{}
(OMT-based) with/out sequential-counter encoding (left) and with/out
cardinality-network encoding (right). 
(\textcolor{brown}{Brown} points denote unsatisfiable benchmarks,
\blue{blue}
denote satisfiable ones and \green{green} ones represent timeouts.)
\newline
[Bottom, tables]
Effect of splitting the PB sums into chunks of maximum variable number
(no split, 10, 15, 20 variables)  with the sequential-counter encoding
(left) and the cardinality-network encoding (right).  
% Top: Results on \maxsmt problems using the \omt encoding in \cite{st_tocl14} (top),
% paired with the Boolean cardinality constraint encoding in \cite{Sinz05} (middle) and in \cite{parametricCardinalityConstraint} 
% (bottom). Bottom: performance gain obtained by joining the input formula with the sequential counter circuit (left)
% and the cardinality network encoding (right); brown colour denotes unsatisfiable benchmarks, blue represent satisfiable ones and green represents timeouts.
}
% \caption{
% \textmd{Top: the performance gain obained by the limiting the size of the generated sorting circuit. Bottom: the same comparison
% as in figure \ref{tab:maxsmt1}, using a circuit size limit of $20$; 
% brown colour denotes unsatisfiable benchmarks, blue represent satisfiable ones and green represents timeouts.
% }} 
%\end{subfigure}
\end{figure}

%% file: expeval_puremaxsmt.tex
%% TODO: new title?

% Another variant of experiment $\#1$ \RSTODO{fix reference if necessary} that we considered, was to
% transform the original Partial Weighted \maxsmt into Partial \maxsmt by
% setting every weight to be $1$.

%\begin{rschange}
Our second experiment is a variant of the previous one,
in which we consider only single-objective optimizations and we
fix all weights to $1$, so that each problem is encoded as a plain
un-weighted \maxsmt problem. We set the timeout to $100s$. 
The results are reported in Figure~\ref{tab:w1}.

\input{figure-puremaxsmt}

{As far as \optimathsat{} (\omt-based) is concerned,} 
extending the input formula with either of the sorting 
networks increases the number of benchmarks solved within the
timeout. 
%\PTTODO{Check!} 
Surprisingly, this time the sequential counter network 
performs significantly better than the cardinality network, despite
its bigger size. (We do not have a clear explanation of this fact.)

As far as \optimathsat{} (\maxsat-based) is concerned, 
we notice that it significantly outperforms all \omt-based techniques,
solving all problems.
Extending the input formula with the cardinality networks 
slightly worsens the performances.
Also the lemma-lifting techniques outperforms all \omt-based
techniques, solving only two problem less than the previous
\maxsat-based techniques.

As far as \nuz{} (\maxsat-based) is concerned, 
we notice that using the max-resolution \maxsat algorithm it is the
best scorer, although the differences wrt. \optimathsat{}
(\maxsat-based) are  negligible, 
whilst by using the \texttt{wmax}
engine the performances decrease
drastically.

%% file: figure-puremaxsmt.tex
\begin{figure}[t]
\centering
\begin{tabular}{|l|rrr|r|}
\hline
\textbf{tool, configuration \& encoding} & \textbf{inst.} & \textbf{term.} & \textbf{timeout} & \textbf{time (s.)} \\
\hline
%\multicolumn{5}{|c|}{\optimathsat} \\
\optimathsat (\omt-based)   	&      2499	    &  1794     &  705	        &    11178         \\
\optimathsat (\omt-based + seq. counter)	 	&      2499	    &  {\bf 2451}     &  48	        &    18033         \\
\optimathsat (\omt-based + card. network) 	&      2499	    &  2186     &  313	        &    10633         \\
\hline
%\multicolumn{5}{|c|}{\optimathsat + maxres} \\
 \optimathsat (\maxsat-based w. maxres)	&      2499	    &  {\bf 2499}     &    0	        &      {\bf 128}         \\
 \optimathsat (\maxsat-based w. maxres + seq. counter)		&      2499	    &  2499     &    0	        &     1638         \\
 \optimathsat (\maxsat-based w. maxres + card. netw.)		&      2499	    &  2499     &    0	        &      257         \\
\hline
            \optimathsat (lemma-lifting w. \maxino)	&      2499	    &  2497     &    2	        &      343         \\
\hline
\hline
               \vZ (\maxsat-based w. maxres)	&      2499	    &  {\bf \blue{2499}}     &    0	        &      {\bf \blue{119}}         \\
               \vZ (\maxsat-based w. wmax)	&      2499	    &  1799     &  733	        &    10549         \\
\hline
\end{tabular}
% NOTE: I don't think any of the available plots is significant
%\vspace{0.5em}
%\begin{tabular}{cc}%
%\hspace{-2.5em}
%\includegraphics[scale=0.36]{png/}
%&
%\hspace{-2.0em}
%\includegraphics[scale=0.36]{png/}\\
%\end{tabular}
\caption{\label{tab:w1}\textmd{
Results of various solvers, configurations and encodings
on  CGM-encoding  problems of  \cite{nguyensgm_er16,nguyensgm16} with 
single-objective weight-$1$.
\ignoreinshort{(Values in \textbf{boldface} denote the best performance of each
category; values in \textbf{\blue{blue}} denote the absolute best performance.)}
}}
\end{figure}

%% file: unsuitable.tex
Here we present a couple of test sets which cannot be supported 
by any \maxsat-based technique in \optimathsat or \nuz and,
to the best of our knowledge, no encoding of these
problem into \maxsmt 
has ever been conceived.
Thus the solution is restricted to \omt-based
techniques. (To this extent, with \optimathsat we have used the
linear-search strategy rather than the default adaptive linear/binary
one to better compare with the linear strategy adopted by \nuZ.)

%% file: expeval_maxmin.tex
% NOTA: questo file è fatto un po' alla svelta, per usare un eufemismo, perche'
% sono imbarcato con diverse cose da concludere al più presto, scusa.

% NOTA 2: ho lasciato le reference come TODO nel caso in cui l'ordine degli esperimenti cambi

% \subsection*{Benchmark Set \#3: Optimal Realization of Goal Model with Soft-Requirements}
%\label{exp:mm}
%% WARNING: labels do not work on subsections which number is hidden (with * parameter)
%% 

% In this performance evaluation, we derived a new benchmark-set from the multi-objective 
% formulas used in experiment $\#1$ \RSTODO{fix ref if necessary}. More in detail, the lexicographic combination 
% of the \maxsmt goals is replaced with a \maxmin combination, in which each \maxsmt 
% objective $\cost$ is normalised to the interval $[0, 1]$ using as rough upper-bound 
% the sum of weights of all soft-constraints.

% The goal of a \maxmin problem is to find the minimum value of $\cost{}$ s.t.
% $\bigwedge_{j=0}^{3} (\cost_{j} \leq \cost) \wedge \bigvee_{i=0}^{N} (\cost_{i} = \cost{})$,
% $\cost$ being a fresh objective variable.

%% TODO: a cosa corrisponde fare maxmin in un scenario reale?

%%%% max-min PLOTS & TABLE
\input{figure-maxmin.tex}

%\RSTODO{giustificare lilnear}
%\RSTODO{dire perche' scatterplots anche con \nuz.}
%\begin{rschange}
In our third experiment we consider another variant of the problems in
Test Set \#1, in which the three PB/\maxsmt{} objectives
\tuple{\cost_1,\cost_2,\cost_3} are subject to a  \maxmin combination:
each objective $\cost_j$ is normalized so that its range equals
 $[0, 1]$ (i.e., it is divided by $\sum_i w_{ji}$), then
 $\bigwedge_{j=1}^{3} (\cost_j \leq \cost)$ s.t. \cost{} is a fresh
\larat variable 
is added to the main formula, and the solver is asked to find a 
 solution making \cost minimum (see \cite{st_cav15}).~
{Notice that max-min optimization guarantees a sort of
  ``fairness'' among the objectives $\cost_1,...,\cost_3$.}
%Min-max problems are dual.
%
Since the problem is more complex than the previous ones and the 
most-efficient \maxsat-based techniques are not applicable,
we increased the timeout to $300$ seconds. 
The results are shown in Figure~\ref{tab:mm}. 
(Unlike with Figure~\ref{tab:maxsmt1}, 
since the difference in performance between \optimathsat{} with
 the two sorting networks is minor,
here and in Figure~\ref{tab:teso1} we have dropped the scatterplot with the sequential-counter
encoding and we introduced one comparing with \nuz{} instead.)

%\TODO{RIVEDERE CON I NUOVI RISULTATI!}
Looking at the table and at the scatterplot on the left, 
we notice that enhancing the \omt-based technique of \optimathsat by
adding the cardinality networks improves significantly the
performances. 
Also, looking at the table and  at the scatterplot on
the right, 
we notice that \omt-based technique of \optimathsat,  with the help of
sorting networks, performs
equivalently or slightly 
better than that of \nuz. 
%\end{rschange}

% From the experimental data, shown in figure \ref{tab:mm}, it can be seen that
% extending \optimathsat{} with cardinality networks both increases the number of 
% solved benchmarks and decreases the search time. This performance gain becomes 
% increasingly more significant with harder formulas.
%
% In this experiment, \vZ{} cannot take advantage of its \maxsat engines for dealing
% with our formulas, and falls back on an optimization routine based on 
% \cite{nieuwenhuis_sat06,st-ijcar12,st_tocl14}, like \optimathsat{}. ...
% TODO: verificare she nieuwenuis_sat06 necessario

%% file: figure-maxmin.tex
\begin{figure}[t]
\centering
\begin{tabular}{|l|rrr|r|}
\hline
\textbf{tool, configuration \& encoding} & \textbf{inst.} & \textbf{term.} & \textbf{timeout} & \textbf{time (s.)} \\
\hline
% LINEAR
\optimathsat (\omt-based)                 & 2399 & 2340              &  59 & 20841 \\
\optimathsat (\omt-based + seq. counter)  & 2399 & 2394              &   5 & 9511 \\
\optimathsat (\omt-based + card. network) & 2399 & {\bf \blue{2395}} &   4 & 8275 \\
\hline
\ignore{ %%%% OK TENIAMO LA LINEAR
% ADAPTIVE
\optimathsat (\omt-based)                 & 2399 & 2326               &   73 & 20024 \\
%
% CONTROL GROUP: should have performance comparable to \optimathsat (\omt-based)
% to guarantee fairness of vZ results.
%\optimathsat [same formulas as vZ]       & 2399 & 2336               &   63 & 18534 \\
%
\optimathsat (\omt-based + seq. counter)  & 2399 & 2388               &   11 &  8977 \\
\optimathsat (\omt-based + card. network) & 2399 & {\bf \blue{2392}}  &    7 &  8648 \\
\hline
} %%%%%%%%%%%%%%%%%%%%%%%
\vZ  (\omt-based)                         & 2399 & 2390               &    9 &  8076 \\
\hline
\end{tabular}
\vspace{0.5em}
\begin{tabular}{cc}%
\ignore{ %%%% OK TENIAMO LA LINEAR
\hspace{-2.5em}
\includegraphics[scale=0.36]{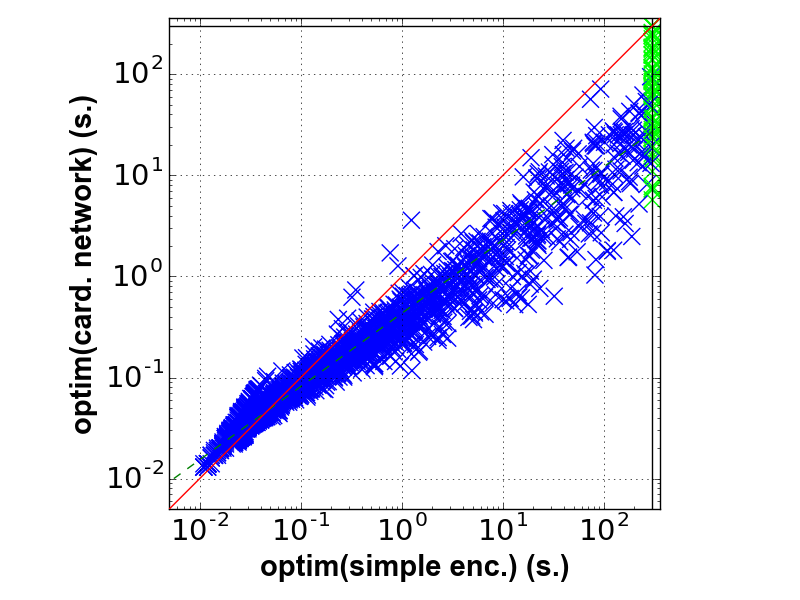}
&
\hspace{-3.0em}
\includegraphics[scale=0.36]{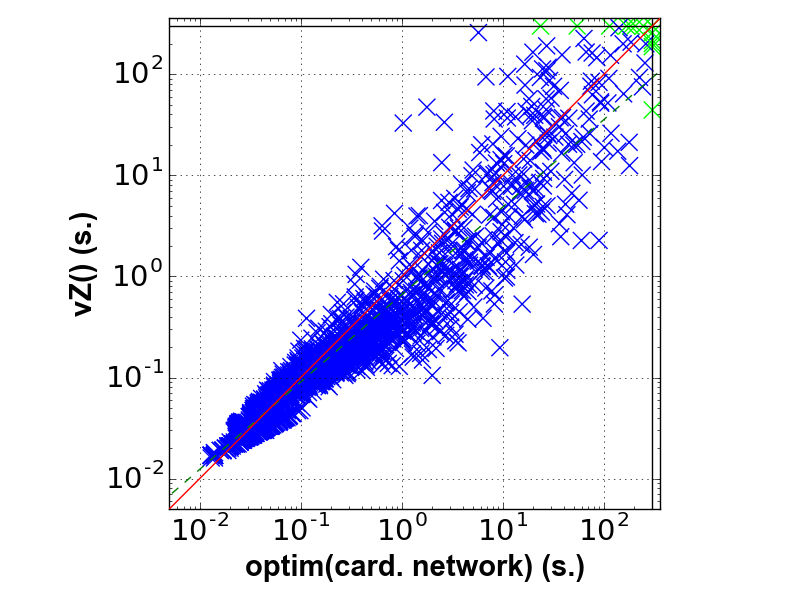}
\\
} %%%%%%%%%%%%%%%%%%%%%%%
\hspace{-2.5em}
\includegraphics[scale=0.36]{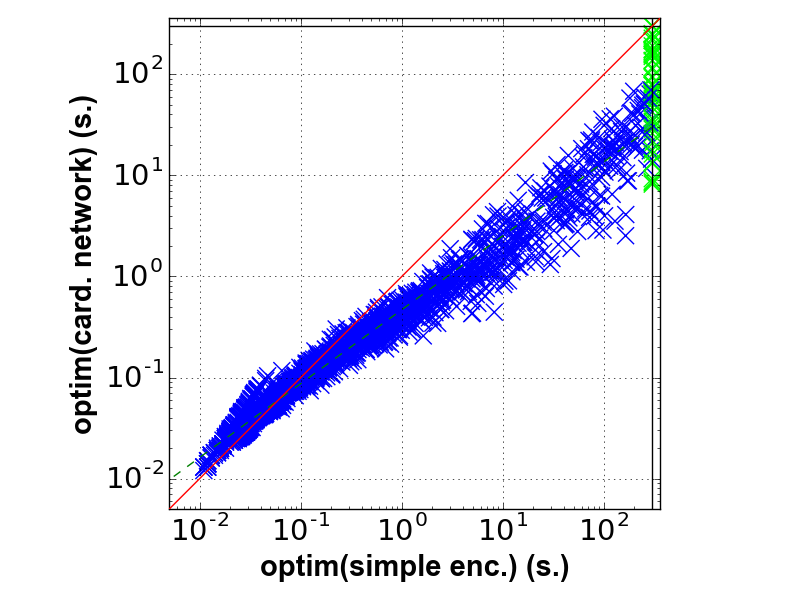}
&
\hspace{-3.0em}
\includegraphics[scale=0.36]{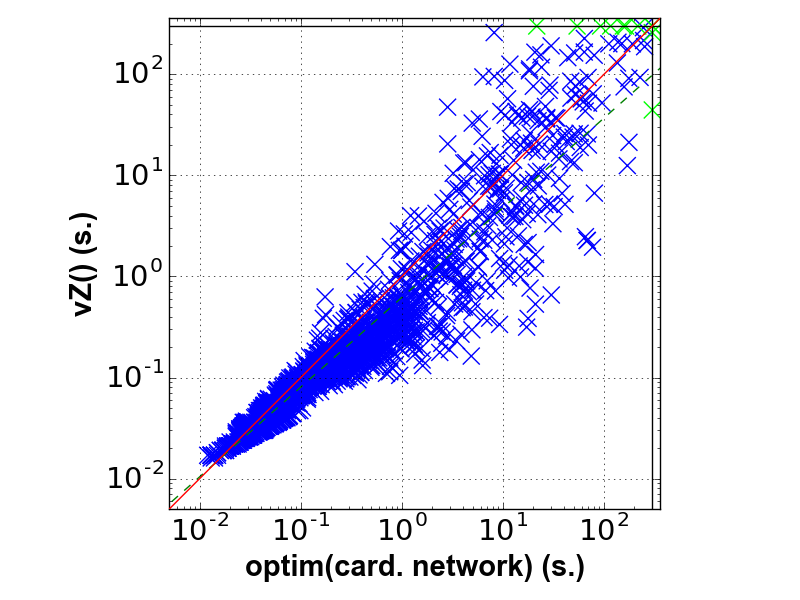}
\\
\end{tabular}
% \begin{tabular}{cc}%
% \hspace{-2.5em}
% \includegraphics[scale=0.36]{png/mm_simple_nlln_lin.png}
% &
% \hspace{-2.0em}
% \includegraphics[scale=0.36]{png/mm_nlln_vz_lin.png}\\
% \end{tabular} 
\caption{\label{tab:mm}
\textmd{[Table:]
results of various solvers with \omt-based  configurations
on  CGM-encoding  problems of  \cite{nguyensgm_er16,nguyensgm16}  
with max-min objective functions.
\newline
[Left scatterplot:] \optimathsat + card. network
vs. plain \optimathsat.
\newline
[Right scatterplot:] \nuz vs. \optimathsat  + card. network.
\ignore{
Top: comparison among \optimathsat and \vZ on a set of benchmarks with
a \maxmin combination of \maxsmt goals. Bottom-left: improvement gained by
using cardinality networks in \optimathsat{}. Bottom-right: comparison among
the best configuration of \optimathsat and \vZ.
}}}
\end{figure}

%% file: expeval_mixed_lmt.tex
%\subsection*{Benchmark Set \#2: Structured Learning Modulo Theories}

%\begin{rschange}
  
\input{figure-mixedlmt.tex}

In our  fourth experiment we consider a set of $500$ problems taken from
\textsc{PyLMT} \cite{pylmt_url}, a tool for Structured Learning Modulo
Theories \cite{teso_aij15} which uses \optimathsat as back-end oracle
for performing inference in the context of machine learning in hybrid
domains.
\ignore{Starting from the original set of $500$ satisfiable formulas, which
used a naive encoding of Pseudo-Boolean objectives by default, we
generated a new benchmark set which uses the \texttt{assert-soft}
statement to encode the following objective function given by a
non-trivial combination of several Pseudo-Boolean terms.}
The objective functions $\cost$  are complex combinations of PB
functions in the form:
\begin{eqnarray}
\cost &\defas&
\textstyle
\sum_j w_j \cdot B_j + cover - \sum_k w_k \cdot C_k - | K - cover |,
\\
s.t.\ cover &\defas&
\textstyle
 \sum_i w_i A_i ,
\ignore{
\\
\cost \defas
\textstyle
\sum_j v_j \cdot B_j + cover - \sum_k z_k \cdot C_k - | K - cover |,
&\ \  &
\textstyle
cover \defas
 \sum_i w_i A_i, 
}
\end{eqnarray}
$A_i,B_j,C_k$ being Boolean atoms, $w_i,v_j,z_k,K$ being rational
constants. 
\ignore{As with the previous case, 
no \maxsat-based technique in \optimathsat or \nuz supports this kind
of problems and, to the best of our knowledge, no encoding of this
problem into \maxsmt has ever been conceived, so that the solution is restricted to \omt-based
techniques.
}
We imposed a timeout of $600$ seconds.
The results are presented in Figure~\ref{tab:teso1}.
% for both tools, and verified that all 
%solvers and their configurations --when terminating-- agreed on the optimal solution values.

Looking at the table and at the scatterplot on the left, 
we notice that enhancing the \omt-based technique of \optimathsat by
adding the cardinality networks improves the
performances, although this time the improvement is not dramatic. 
(We believe this is due that the values of the weights $w_i,v_j,z_k,K$
are very heterogeneous, not many weights share the same value.)
Also, looking at the table and  at the scatterplot on
the right, 
we notice that \omt-based technique of \optimathsat performs significantly
better than that of \nuz, even without the help of sorting networks. 

\ignore{
\TODO{LASCIARE QUESTO PARAGRAFO?} Finally, as a matter of fairness wrt. \nuz, 
we wondered if the test was affected by the usage of {\tt
  assert-soft} construct (which could not be used in \nuz). 
To address this fact, we fed to \optimathsat{} the same encoding of \nuz, without {\tt
  assert-soft}, an displayed the results in Figure~\ref{tab:teso1}.
We notice that with \optimathsat (\omt-based) not only the usage of  {\tt
  assert-soft} provides no advantage, but also it slightly worsens the
performances. 
}

\ignore{
We ran both \optimathsat and \nuz (\omt-based)  over the original set of benchmarks,
and compared the following three configurations for \optimathsat over
the \texttt{assert-soft} based benchmark set: \textit{simple enc.},
which maps the input weighted \maxsmt formula into an \omt problem
with no circuit, \textit{seq. counter enc.}, which extends the input
formula with the sequential counter encoding, and
\textit{card. network enc.} which uses instead the cardinality network
encoding.
}

%\end{rschange}

\ignore{
\RSTODO{spiega} The results, depicted in figure \ref{tab:teso1}, show
that extending the input problems with sorting network circuits
benefits the number of solved benchmarks within the timeout, with the
cardinality network encoding performing better by a small margin.  As
highlighted by the scatter plots in the bottom part of figure
\ref{tab:teso1}, generating these circuit on the fly causes a limited
overhead on \RSCHANGE{very-}easy benchmarks, whereas it can
significantly improve the running time on difficult ones.
}

%% file: figure-mixedlmt.tex
\begin{figure}[tb]
\centering
\begin{tabular}{|l|rrr|r|}
\hline
\textbf{tool, configuration \& encoding}    & \textbf{inst.} & \textbf{term.} & \textbf{timeout} & \textbf{time (s.)} \\
% \hline
% \multicolumn{5}{|c|}{original encoding} \\
% \multicolumn{5}{|c|}{\optimathsat using assert-soft} \\
\hline
\optimathsat (\omt-based) & 500 & 421 & 79 & 2607 \\
\optimathsat (\omt-based + seq. counter) & 500 & 441 & 59 & 6381 \\
\optimathsat (\omt-based + card. network) & 500 & \textbf{\blue{442}} & 58 & 6189 \\
\hline
\hline
\vZ (\omt-based) & 500 & 406 & 94 & 2120 \\
% \hline
% \optimathsat (\omt-based, no assert-soft) & 500 & 424 & 76 & 3522 \\
\hline
\end{tabular}

\vspace{0.5em}

\begin{tabular}{cc}%
\ignore{
\hspace{-2.5em}
		\includegraphics[scale=0.4]{png/ml_s_vz.png}
&
\hspace{-3em}
		\includegraphics[scale=0.4]{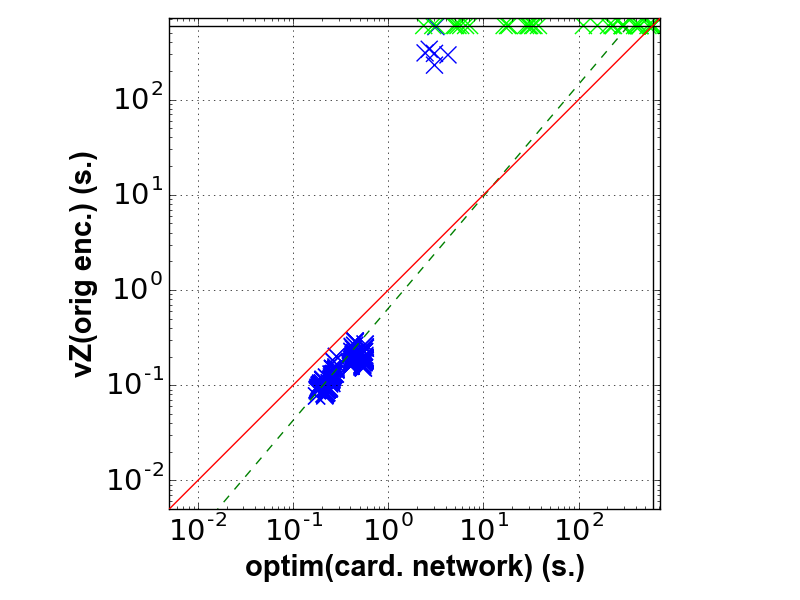}\\
\hspace{-2.5em}
		\includegraphics[scale=0.35]{png/ml_s_nn.png}
&
\hspace{-2em}
		\includegraphics[scale=0.35]{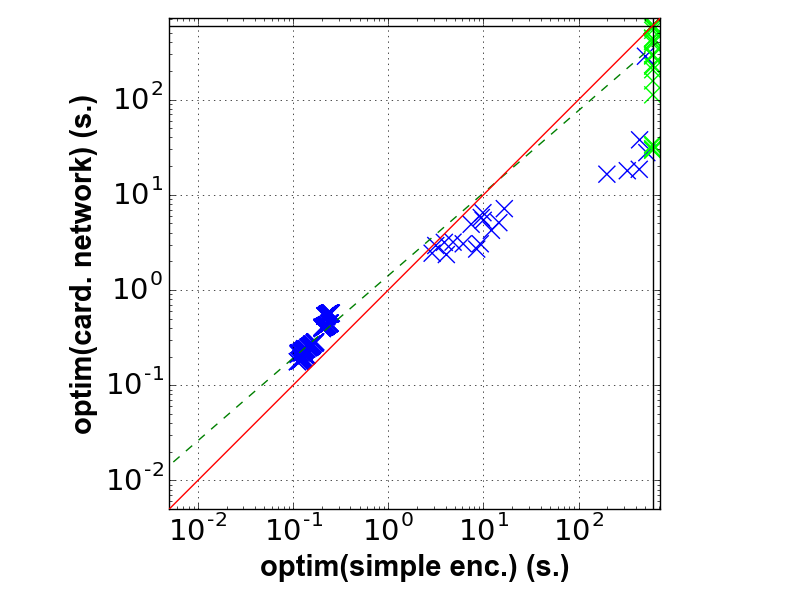}
\\
}
\hspace{-2.5em}
		\includegraphics[scale=0.35]{png/ml_s_nlln.png}
\hspace{-2.5em}
		\includegraphics[scale=0.35]{png/ml_nlln_vz.png}
\end{tabular}

\caption{\label{tab:teso1}
\ignore{\textmd{Top: comparison among various \optimathsat configurations and \vZ on a set of benchmarks 
based on the Structured Learning Modulo Theories domain \cite{teso_aij15}. Bottom: scatter plots
for the same experimental data showing a comparison among \optimathsat solve time over the 
original problems wrt. the \texttt{assert-soft} encoding enriched with a sorting network circuit
generated on the fly.}}
\textmd{\textmd{[Table:]
results of various solvers with \omt-based  configurations
on LMT-encoding  problems of  \cite{teso_aij15}  
with complex objective functions.
\newline
[Left scatterplot:] \optimathsat + card. network
vs. plain \optimathsat.
\newline
[Right scatterplot:] \nuz vs. \optimathsat  + card. network.
}
}}
\end{figure}

%% file: expeval_discussion.tex
%\RSTODO{BLA BLA (vedi intro)}

% This shows that using a dedicated \maxsat engine can outperform state-of-the-art \omt 
% optimization techniques, although with the limitation of being less expressive in terms
% of freedom of combining objectives.

We summarize the results as follows.
\begin{aenumerate}
\item When applicable, \omt-based approaches performed much  better than 
\omt-based ones, in particular when adopting Maximum-Resolution 
 as \maxsat{} engine.

% Comparing  \maxsat-based approaches wrt. \omt-based ones on 
%  problems where the former are applicable, it turns out that the former provide better performances, in particular when adopting Maximum-Resolution 
% \cite{narodytskab14,bjorner_scss14} as \maxsat{} engine. 
\item Bidirectional 
sorting-network encodings improved significantly the performances 
of \omt-based approaches, and often also of \maxsat-based ones.
\item
\nuz performed better than \optimathsat{} 
on \maxsat{}-based approaches,
whilst the latter performed sometimes similarly and sometimes significantly 
better on \omt-based ones, in particular when enhanced
 by the sorting-network encodings. 
\end{aenumerate}

%% file: concl.tex
%\RSTODO{RISCRIVERE}
\maxsmt and OMT with Pseudo-Boolean objective functions 
are important sub-cases of OMT,
 for which specialized techniques have
been developed over the years, in particular exploiting
state-of-the-art \maxsat procedures. 
When applicable, these specialized procedures seem to be more efficient
than general-purpose OMT. 
When they are not applicable, 
OMT-based technique can strongly benefit from the integration with
bidirectional sorting networks to deal with PB components of
objectives. 

OMT is a young technology, with large margins for
improvements. Among others, one interesting research
direction is that of integrating \maxsat-based techniques with
standard \omt-based ones for efficiently  handling complex objectives and
constraints, so that to combine the efficiency of the former with the
expressivity of the latter.

%% file: main.bbl
\begin{thebibliography}{10}

\bibitem{tacas17-experiments-url}
\url{http://disi.unitn.it/trentin/resources/tacas17.tar.gz}.

\bibitem{cgm-tool}
{CGM-Tool}.
\newblock \url{http://www.cgm-tool.eu}.

\bibitem{nuz-url}
{$\mu$Z}.
\newblock \url{http://rise4fun.com/z3opt}.

\bibitem{optimathsat-url}
{OptiMathSAT}.
\newblock \url{http://optimathsat.disi.unitn.it}.

\bibitem{pylmt_url}
{PyLMT}.
\newblock \url{http://www.bitbucket.org/stefanoteso/pylmt}.

\bibitem{parametricCardinalityConstraint}
I.~{Ab{\'\i}o}, R.~{Nieuwenhuis}, A.~{Oliveras}, and
  E.~{Rodr\'{\i}guez-Carbonell}.
\newblock {A Parametric Approach for Smaller and Better Encodings of
  Cardinality Constraints}.
\newblock In {\em 19th International Conference on Principles and Practice of
  Constraint Programming}, CP'13, 2013.

\bibitem{alviano_ijcai15}
M.~Alviano, C.~Dodaro, and F.~Ricca.
\newblock A maxsat algorithm using cardinality constraints of bounded size.
\newblock In {\em Proceedings of the 24th International Conference on
  Artificial Intelligence}, IJCAI'15, pages 2677--2683. AAAI Press, 2015.

\bibitem{AnsoteguiBPSV11}
C.~Ans{\'o}tegui, M.~Bofill, M.~Palah\'{\i}, J.~Suy, and M.~Villaret.
\newblock {Satisfiability Modulo Theories: An Efficient Approach for the
  Resource-Constrained Project Scheduling Problem}.
\newblock In {\em SARA}, 2011.

\bibitem{Asin2011}
R.~As{\'i}n, R.~Nieuwenhuis, A.~Oliveras, and E.~Rodr{\'i}guez-Carbonell.
\newblock Cardinality networks: a theoretical and empirical study.
\newblock {\em Constraints}, 16(2):195--221, 2011.

\bibitem{BSST09HBSAT}
C.~Barrett, R.~Sebastiani, S.~A. Seshia, and C.~Tinelli.
\newblock {\em Satisfiability Modulo Theories}, chapter~26, pages 825--885.
\newblock In Biere et~al. \cite{HandbookOfSAT2009}, February 2009.

\bibitem{HandbookOfSAT2009}
A.~Biere, M.~J.~H. Heule, H.~van Maaren, and T.~Walsh, editors.
\newblock {\em Handbook of Satisfiability}.
\newblock IOS Press, February 2009.

\bibitem{bjorner_private16}
N.~Bjorner.
\newblock personal communication, 02 2016.

\bibitem{bjorner_scss14}
N.~Bjorner and A.-D. Phan.
\newblock {$\nu{}Z$ - Maximal Satisfaction with Z3}.
\newblock In {\em Proc International Symposium on Symbolic Computation in
  Software Science}, Gammart, Tunisia, December 2014. EasyChair Proceedings in
  Computing (EPiC).
\newblock \url{http://www.easychair.org/publications/?page=862275542}.

\bibitem{bjorner_tacas15}
N.~Bjorner, A.-D. Phan, and L.~Fleckenstein.
\newblock {\nuz{} - An Optimizing \smt{} Solver}.
\newblock In {\em Proc. TACAS}, volume 9035 of {\em LNCS}. Springer, 2015.

\bibitem{cimattifgss10}
A.~Cimatti, A.~Franz{\'e}n, A.~Griggio, R.~Sebastiani, and C.~Stenico.
\newblock Satisfiability modulo the theory of costs: Foundations and
  applications.
\newblock In {\em TACAS}, volume 6015 of {\em LNCS}, pages 99--113. Springer,
  2010.

\bibitem{cgss_sat13_maxsmt}
A.~Cimatti, A.~Griggio, B.~J. Schaafsma, and R.~Sebastiani.
\newblock {A Modular Approach to MaxSAT Modulo Theories}.
\newblock In {\em International Conference on Theory and Applications of
  Satisfiability Testing, SAT}, volume 7962 of {\em LNCS}, July 2013.

\bibitem{EenS06}
N.~E{\'{e}}n and N.~S{\"{o}}rensson.
\newblock Translating pseudo-boolean constraints into {SAT}.
\newblock {\em {JSAT}}, 2(1-4):1--26, 2006.

\bibitem{larrazorr14}
D.~Larraz, A.~Oliveras, E.~Rodr{\'{\i}}guez{-}Carbonell, and A.~Rubio.
\newblock {Minimal-Model-Guided Approaches to Solving Polynomial Constraints
  and Extensions}.
\newblock In {\em {SAT}}, 2014.

\bibitem{li_popl14}
Y.~Li, A.~Albarghouthi, Z.~Kincad, A.~Gurfinkel, and M.~Chechik.
\newblock {Symbolic Optimization with SMT Solvers}.
\newblock In {\em POPL}, 2014.

\bibitem{MSLM09HBSAT}
J.~P. Marques-Silva, I.~Lynce, and S.~Malik.
\newblock {\em Conflict-Driven Clause Learning SAT Solvers}, chapter~4, pages
  131--153.
\newblock In Biere et~al. \cite{HandbookOfSAT2009}, February 2009.

\bibitem{NadelR16}
A.~Nadel and V.~Ryvchin.
\newblock Bit-vector optimization.
\newblock In {\em Tools and Algorithms for the Construction and Analysis of
  Systems, {TACAS} 2016}, volume 9636 of {\em LNCS}. Springer, 2016.

\bibitem{narodytskab14}
N.~Narodytska and F.~Bacchus.
\newblock Maximum satisfiability using core-guided maxsat resolution.
\newblock In {\em Proceedings of the Twenty-Eighth {AAAI} Conference on
  Artificial Intelligence, July 27 -31, 2014, Qu{\'{e}}bec City, Qu{\'{e}}bec,
  Canada.}, pages 2717--2723. {AAAI} Press, 2014.

\bibitem{nguyensgm16}
C.~M. Nguyen, R.~Sebastiani, P.~Giorgini, and J.~Mylopoulos.
\newblock Multi object reasoning with constrained goal models.
\newblock {\em Requirement Engineering}, 2016.
\newblock To appear.

\bibitem{nguyensgm_er16}
C.~M. Nguyen, R.~Sebastiani, P.~Giorgini, and J.~Mylopoulos.
\newblock {Requirements Evolution and Evolution Requirements with Constrained
  Goal Models}.
\newblock In {\em Proceedings of the 37nd International Conference on
  Conceptual Modeling - ER16}, LNCS. Springer, 2016.

\bibitem{nieuwenhuis_sat06}
R.~Nieuwenhuis and A.~Oliveras.
\newblock {On SAT Modulo Theories and Optimization Problems}.
\newblock In {\em Proc. Theory and Applications of Satisfiability Testing - SAT
  2006}, volume 4121 of {\em LNCS}. Springer, 2006.

\bibitem{RM09HBSAT}
O.~Roussel and V.~Manquinho.
\newblock {\em Pseudo-Boolean and Cardinality Constraints}, chapter~22, pages
  695--733.
\newblock In Biere et~al. \cite{HandbookOfSAT2009}, February 2009.

\bibitem{sawayag05}
N.~W. Sawaya and I.~E. Grossmann.
\newblock A cutting plane method for solving linear generalized disjunctive
  programming problems.
\newblock {\em Computing Chemical Engineering}, 29(9):1891--1913, 2005.

\bibitem{sebastiani07}
R.~Sebastiani.
\newblock {Lazy Satisfiability Modulo Theories}.
\newblock {\em Journal on Satisfiability, Boolean Modeling and Computation,
  JSAT}, 3(3-4):141--224, 2007.

\bibitem{st-ijcar12}
R.~Sebastiani and S.~Tomasi.
\newblock {Optimization in SMT with LA(Q) Cost Functions}.
\newblock In {\em IJCAR}, volume 7364 of {\em LNAI}, pages 484--498. Springer,
  July 2012.

\bibitem{st_tocl14}
R.~Sebastiani and S.~Tomasi.
\newblock {Optimization Modulo Theories with Linear Rational Costs}.
\newblock {\em ACM Transactions on Computational Logics}, 16(2), March 2015.

\bibitem{st_cav15}
R.~Sebastiani and P.~Trentin.
\newblock {OptiMathSAT: A Tool for Optimization Modulo Theories.}
\newblock In {\em Proc. International Conference on Computer-Aided
  Verification, CAV 2015}, volume 9206 of {\em LNCS}. Springer, 2015.

\bibitem{st_tacas15}
R.~Sebastiani and P.~Trentin.
\newblock {Pushing the Envelope of Optimization Modulo Theories with
  Linear-Arithmetic Cost Functions}.
\newblock In {\em Proc. Int. Conference on Tools and Algorithms for the
  Construction and Analysis of Systems, TACAS'15}, volume 9035 of {\em LNCS}.
  Springer, 2015.

\bibitem{st_smt16}
R.~Sebastiani and P.~Trentin.
\newblock O{n the Benefits of Enhancing Optimization Modulo Theories with
  Sorting Networks for MaxSMT}.
\newblock In {\em Proceedings of the 14th International Workshop on
  Satisfiability Modulo Theories, SMT-2016.}, {CEUR} Workshop Proceedings,
  2016.

\bibitem{Sinz05}
C.~Sinz.
\newblock Towards an optimal cnf encoding of boolean cardinality constraints.
\newblock In P.~van Beek, editor, {\em CP}, volume 3709 of {\em LNCS}, pages
  827--831. Springer, 2005.

\bibitem{teso_aij15}
S.~Teso, R.~Sebastiani, and A.~Passerini.
\newblock {Structured Learning Modulo Theories.}
\newblock {\em Artificial Intelligence Journal}, 2015.
\newblock In print. Available online 29 April 2015.
  \url{http://dx.doi.org/10.1016/j.artint.2015.04.002.}

\end{thebibliography}
